\documentclass[12pt,a4paper]{article}
\usepackage{graphics,graphicx}
\usepackage{amssymb,epsfig,amsmath,euscript,array}
\usepackage{cite}

\makeatletter
\@addtoreset{equation}{section}
\makeatother
\renewcommand{\theequation}{\thesection.\arabic{equation}}

\newcommand{\startappendix}{
\setcounter{section}{0}
\renewcommand{\thesection}{\Alph{section}}
\renewcommand{\theequation}{\Alph{section}.\arabic{equation}}}

\newcommand{\Appendix}[1]{
\refstepcounter{section}
\begin{flushleft}
{\Large\bf Appendix \thesection: #1}
\end{flushleft}}

\newcounter{multieqs}

\newenvironment{pretty}{}{}

\newcommand{\be}{\begin{equation}}
\newcommand{\ee}{\end{equation}}

\newcommand{\bm}[1]{\mbox{\boldmath $#1$}}

\newcommand{\kslash}{k \!\!\! / }
\newcommand{\qslash}{q \!\!\! / }
\newcommand{\lslash}{l \!\! / }
\newcommand{\Pslash}{P \!\!\!\! / }
\newcommand{\Qslash}{Q \!\!\!\! / }
\newcommand{\islash}{i \!\!\! / }
\newcommand{\jslash}{j \!\!\! / }
\newcommand{\aslash}{a \!\!\! / }
\newcommand{\bslash}{{b \hspace{-6pt} \slash} }

\newcommand{\onslash}{1 \!\!\! / }
\newcommand{\twslash}{2 \!\!\!/ }
\newcommand{\thslash}{3 \!\!\!/ }
\newcommand{\foslash}{4 \!\!\! / }
\newcommand{\fislash}{5 \!\!\! / }

\newcommand{\mslash}{m \!\!\! / }

\def\bd{\begin{document}}
\def\ed{\end{document}}
\def\nn{\nonumber}
\def\bea{\begin{eqnarray}}
\def\eea{\end{eqnarray}}

\def\ab{(ijab)}
\def\ba{(ijba)}
\def\ijab{{\tr}_{+}(\islash\, \jslash\, \aslash \, \bslash)}
\def\ijba{{\tr}_{+}(\islash\, \jslash\, \bslash \, \aslash)}
\def\ijaP{{\tr}_{+}(\islash\, \jslash\, \aslash \, \Pslash)}
\def\ijPLa{{\tr}_{+}(\islash\, \jslash\, \Pslash_L \, \aslash)}
\def\ijaPL{{\tr}_{+}(\islash\, \jslash\, \aslash \, \Pslash_L)}
\def\ijPLza{{\tr}_{+}(\islash\, \jslash\, \Pslash_{L;z} \, \aslash)}
\def\ijaPLz{{\tr}_{+}(\islash\, \jslash\, \aslash \, \Pslash_{L;z})}
\def\ijPa{{\tr}_{+}(\islash\, \jslash\, \Pslash \, \aslash)}
\def\iaPb{{\tr}_{+}(\islash\, \aslash\, \Pslash \, \bslash)}
\def\ibPa{{\tr}_{+}(\islash\, \bslash\, \Pslash \, \aslash)}
\def\ijPmu{{\tr}_{+}(\islash\, \jslash\, \Pslash \, \mu)}
\def\ibmuP{{\tr}_{+}(\islash\, \bslash\, \mu \, \Pslash)}
\def\ibmua{{\tr}_{+}(\islash\, \bslash\, \mu \, \aslash)}
\def\iamub{{\tr}_{+}(\islash\, \aslash\, \mu \, \bslash)}
\def\jaPb{{\tr}_{+}(\jslash\, \aslash\, \Pslash \, \bslash)}
\def\ijmuP{{\tr}_{+}(\islash\, \jslash\, \mu \, \Pslash)}
\def\ijmum{{\tr}_{+}(\islash\, \jslash\, \mu \, \mslash)}
\def\ijmmu{{\tr}_{+}(\islash\, \jslash\, \mslash \, \mu)}
\def\ijmP{{\tr}_{+}(\islash\, \jslash\, \mslash \, \Pslash)}
\def\iabP{{\tr}_{+}(\islash\, \aslash\, \bslash \, \Pslash)}
\def\ijbP{{\tr}_{+}(\islash\, \jslash\, \bslash \, \Pslash)}
\def\jbPa{{\tr}_{+}(\jslash\, \bslash\, \Pslash \, \aslash)}
\def\ijPb{{\tr}_{+}(\islash\, \jslash\, \Pslash \, \bslash)}
\def\jbmua{{\tr}_{+}(\jslash\, \bslash\, \mu \, \aslash)}

\def\loablt{ {\tr}_{+}(\lslash_1\, \aslash \, \bslash\, \lslash_2)}

\def\ijlolt{{\tr}_{+}(\islash\, \jslash\, \lslash_1 \, \lslash_2)}
\def\ijltlo{{\tr}_{+}(\islash\, \jslash\, \lslash_2 \, \lslash_1)}
\def\ibloa{{\tr}_{+}(\islash\, \bslash\, \lslash_1 \, \aslash)}
\def\jaltb{{\tr}_{+}(\jslash\, \aslash\, \lslash_2 \, \bslash)}
\def\ialtb{{\tr}_{+}(\islash\, \aslash\, \lslash_2 \, \bslash)}
\def\bltloa{{\tr}_{+}(\bslash\, \lslash_2\, \lslash_1 \, \aslash)}
\def\jbloa{{\tr}_{+}(\jslash\, \bslash\, \lslash_1 \, \aslash)}
\def\ibPb{{\tr}_{+}(\islash\, \bslash\, \Pslash \, \bslash)}
\def\ijltb{{\tr}_{+}(\islash\, \jslash\, \lslash_2 \, \bslash)}

\def\ijloa{{\tr}_{+}(\islash\, \jslash\,  \lslash_1 \, \aslash)}
\def\ijblt{{\tr}_{+}(\islash\, \jslash\,  \bslash \, \lslash_2)}

\def\jakb{{\tr}_{+}(\jslash\, \aslash\, \kslash \, \bslash)}
\def\iakb{{\tr}_{+}(\islash\, \aslash\, \kslash \, \bslash)}

\def\tofo{{\tr}_{+}(\onslash\, \thslash\, \twslash \, \foslash)}
\def\foto{{\tr}_{+}(\onslash\, \thslash\, \foslash \, \twslash)}
\def\tofi{{\tr}_{+}(\onslash\, \thslash\, \twslash \, \fislash)}
\def\fito{{\tr}_{+}(\onslash\, \thslash\, \fislash \, \twslash)}

\def\lrangle#1#2{\langle #1\,#2\rangle}

\def\Li{{$\rm Li}_2$}

\let\bm=\bibitem
\let\la=\label

\def\npb#1#2#3{Nucl. Phys. {\bf{B#1}} #3 (#2)}
\def\plb#1#2#3{Phys. Lett. {\bf{#1B}} #3 (#2)}
\def\prl#1#2#3{Phys. Rev. Lett. {\bf{#1}} #3 (#2)}
\def\prd#1#2#3{Phys. Rev. {D \bf{#1}} #3 (#2)}
\def\cmp#1#2#3{Comm. Math. Phys. {\bf{#1}} #3 (#2)}
\def\cqg#1#2#3{Class. Quantum Grav. {\bf{#1}} #3 (#2)}
\def\nppsa#1#2#3{Nucl. Phys. B (Proc. Suppl.) {\bf{#1A}}#3 (#2)}
\def\ap#1#2#3{Ann. of Phys. {\bf{#1}} #3 (#2)}
\def\ijmp#1#2#3{Int. J. Mod. Phys. {\bf{A#1}} #3 (#2)}
\def\rmp#1#2#3{Rev. Mod. Phys. {\bf{#1}} #3 (#2)}
\def\mpla#1#2#3{Mod. Phys. Lett. {\bf A#1} #3 (#2)}
\def\jhep#1#2#3{J. High Energy Phys. {\bf #1} #3 (#2)}
\def\atmp#1#2#3{Adv. Theor. Math. Phys. {\bf #1} #3 (#2)}
\newcommand{\EQ}[1]{\begin{equation} #1 \end{equation}}
\newcommand{\AL}[1]{\begin{subequations}\begin{align} #1 \end{align}\end{subequations}}
\newcommand{\SP}[1]{\begin{equation}\begin{split} #1 \end{split}\end{equation}}
\newcommand{\ALAT}[2]{\begin{subequations}\begin{alignat}{#1} #2 \end{alignat}
                        \end{subequations}}
\def\beqa{\begin{eqnarray}}
\def\eeqa{\end{eqnarray}}
\def\beq{\begin{equation}}
\def\eeq{\end{equation}}
\def\sst{\scriptscriptstyle}
\def\thetabar{\bar\theta}
\def\Tr{{\rm Tr}}
\def\one{\mbox{1 \kern-.59em {\rm l}}}
 \def\Nh{\hat{N}}

\def\a{\alpha}      \def\da{{\dot\alpha}}
\def\b{\beta}       \def\db{{\dot\beta}}
\def\c{\gamma}  \def\G{\Gamma}  \def\cdt{\dot\gamma}
\def\d{\delta}  \def\D{\Delta}  \def\ddt{\dot\delta}
\def\e{\epsilon}        \def\vare{\varepsilon}
\def\f{\phi}    \def\F{\Phi}    \def\vvf{\f}
\def\h{\eta}
\def\k{\kappa}
\def\l{\lambda} \def\L{\Lambda}
\def\m{\mu} \def\n{\nu}
\def\o{\omega}
\def\p{\pi} \def\P{\Pi}
\def\r{\rho}
\def\s{\sigma}  \def\S{\Sigma}
\def\t{\tau}
\def\th{\theta} \def\Th{\Theta} \def\vth{\vartheta}
\def\X{\Xeta}
\def\z{\zeta}

\def\cA{{\cal A}} \def\cB{{\cal B}} \def\cC{{\cal C}}
\def\cD{{\cal D}} \def\cE{{\cal E}} \def\cF{{\cal F}}
\def\cG{{\cal G}} \def\cH{{\cal H}} \def\cI{{\cal I}}
\def\cJ{{\cal J}} \def\cK{{\cal K}} \def\cL{{\cal L}}
\def\cM{{\cal M}} \def\cN{{\cal N}} \def\cO{{\cal O}}
\def\cP{{\cal P}} \def\cQ{{\cal Q}} \def\cR{{\cal R}}
\def\cS{{\cal S}} \def\cT{{\cal T}} \def\cU{{\cal U}}
\def\cV{{\cal V}} \def\cW{{\cal W}} \def\cX{{\cal X}}
\def\cY{{\cal Y}} \def\cZ{{\cal Z}}

\def\ua{\underline{\alpha}}
\def\ub{\underline{\phantom{\alpha}}\!\!\!\beta}
\def\uc{\underline{\phantom{\alpha}}\!\!\!\gamma}
\def\um{\underline{\mu}}
\def\ud{\underline\delta}
\def\ue{\underline\epsilon}
\def\una{\underline a}\def\unA{\underline A}
\def\unb{\underline b}\def\unB{\underline B}
\def\unc{\underline c}\def\unC{\underline C}
\def\und{\underline d}\def\unD{\underline D}
\def\une{\underline e}\def\unE{\underline E}
\def\unf{\underline{\phantom{e}}\!\!\!\! f}\def\unF{\underline F}
\def\unm{\underline m}\def\unM{\underline M}
\def\unn{\underline n}\def\unN{\underline N}
\def\unp{\underline{\phantom{a}}\!\!\! p}\def\unP{\underline P}
\def\unq{\underline{\phantom{a}}\!\!\! q}
\def\unQ{\underline{\phantom{A}}\!\!\!\! Q}
\def\unH{\underline{H}}

\def\As {{A \hspace{-6.4pt} \slash}\;}
\def\bs {{b \hspace{-6.4pt} \slash}\;}
\def\Ds {{D \hspace{-6.4pt} \slash}\;}
\def\ds {{\del \hspace{-6.4pt} \slash}\;}
\def\ss {{\s \hspace{-6.4pt} \slash}\;}
\def\ks {{ k \hspace{-6.4pt} \slash}\;}
\def\ps {{p \hspace{-6.4pt} \slash}\;}
\def\pas {{{p_1} \hspace{-6.4pt} \slash}\;}
\def\pbs {{{p_2} \hspace{-6.4pt} \slash}\;}
\def\Ps {{P \hspace{-6.4pt} \slash}\;}
\def\Qs {{Q \hspace{-6.4pt} \slash}\;}

\def\Fh{\hat{F}}
\def\Vh{\hat{V}}
\def\Xh{\hat{X}}
\def\ah{\hat{a}}
\def\xh{\hat{x}}
\def\yh{\hat{y}}
\def\ph{\hat{p}}
\def\xih{\hat{\xi}}
\def\psit{\tilde{\psi}}
\def\Psit{\tilde{\Psi}}
\def\tht{\tilde{\th}}
\def\lt{\tilde{\lambda}}
\def\llt{\tilde{l}}
\def\At{\tilde{A}}
\def\Qt{\tilde{Q}}
\def\Rt{\tilde{R}}
\def\Nt{\tilde{N}}

\def\at{\tilde{a}}
\def\st{\tilde{s}}
\def\ft{\tilde{f}}
\def\pt{\tilde{p}}
\def\qt{\tilde{q}}
\def\vt{\tilde{v}}
\def\nt{\tilde{n}}

\def\delb{\bar{\partial}}
\def\bz{\bar{z}}
\def\bD{\bar{D}}
\def\bB{\bar{B}}

\def\bk{{\bf k}}
\def\bl{{\bf l}}
\def\bp{{\bf p}}
\def\bq{{\bf q}}
\def\br{{\bf r}}
\def\bx{{\bf x}}
\def\by{{\bf y}}
\def\bR{{\bf R}}
\def\bV{{\bf V}}

\def\d{\delta}\def\D{\Delta}\def\ddt{\dot\delta}
\def\pa{\partial} \def\del{\partial}
\def\xx{\times}
\def\uno{\mbox{1 \kern-.59em {\rm l}}}
\def\trp{^{\top}}
\def\inv{^{-1}}
\def\dag{{^{\dagger}}}
\def\pr{^{\prime}}
\def\lan{\langle}
\def\ran{\rangle}
\def\rar{\rightarrow}
\def\lar{\leftarrow}
\def\lrar{\leftrightarrow}
\newcommand{\0}{\,\!}      
\def\one{1\!\!1\,\,}
\def\im{\imath}
\def\jm{\jmath}
\newcommand{\tr}{\mbox{tr}}
\newcommand{\slsh}[1]{/ \!\!\!\! #1}
\def\vac{|0\rangle}
\def\lvac{\langle 0|}
\def\hlf{\frac{1}{2}}
\def\ove#1{\frac{1}{#1}}
\def\Box{\square}
\def\ZZ{\mathbb{Z}}
\def\CC#1{({\bf #1})}
\def\bcomment#1{}
\def\bfhat#1{{\bf \hat{#1}}}
\def\VEV#1{\left\langle #1\right\rangle}
\newcommand{\ex}[1]{{\rm e}^{#1}} \def\ii{{\rm i}}
\def\rr{{\rm r}} \def\rs{{\rm s}}\def\rv{{\rm v}}
\def\ri{{\rm i}}\def\rj{{\rm j}}
\newcommand{\lrbrk}[1]{\left(#1\right)}
\newcommand{\sfrac}[2]{{\textstyle\frac{#1}{#2}}}

\def\Li{{\rm Li}_2}

\font\mybb=msbm10 at 12pt
\def\bb#1{\hbox{\mybb#1}}

\font\myBB=msbm10 at 18pt
\def\BB#1{\hbox{\myBB#1}}

\begin{document}

\begin{flushright}
hep-th/0412108 \\
QMUL-PH-04-13
\end{flushright}

\vspace{20pt}

\begin{center}

{\Large \bf Non-Supersymmetric Loop Amplitudes    }
\\
\vspace{0.3cm}
{\Large \bf  and  MHV Vertices }
\vspace{12pt}
\vspace{33pt}

{\bf James Bedford, Andreas  Brandhuber, Bill  Spence and Gabriele  Travaglini}
\begin{pretty}\footnote{
{\sffamily \{\tt j.a.p.bedford, a.brandhuber, w.j.spence,
g.travaglini\}@qmul.ac.uk }}\end{pretty}

{\em Department of Physics\\
Queen Mary, University of
London\\
Mile End Road, London, E1 4NS\\
United Kingdom
 }

\vspace{40pt} {\bf Abstract}

\end{center}


\noindent
We show how the MHV diagram description of Yang-Mills theories
can be used to study non-supersymmetric loop amplitudes.
In particular, we derive a compact
expression for the cut-constructible part of
the general one-loop MHV multi-gluon
scattering amplitude in pure Yang-Mills theory.
We show that in special cases this expression reduces to
known amplitudes  -- the amplitude with adjacent
negative-helicity gluons, and the five gluon non-adjacent
amplitude. Finally, we briefly
discuss the twistor space interpretation of our result.

\vspace{0.5cm}

\setcounter{page}{0}
\thispagestyle{empty}
\newpage


\section{Introduction}
\setcounter{footnote}{0}
Since Witten's discovery that topological string theory
on super twistor space provides a description of $\cN\!=\!4$
super Yang-Mills  (SYM) \cite{witten}, considerable progress
has been made using twistor-inspired methods to study
Yang-Mills theories.
An important factor in this has been the proposal
to use maximally helicity violating (MHV) diagrams,
built using MHV amplitudes as vertices, in order to derive amplitudes
\cite{csw}.
The MHV diagram construction has the appealing feature that
twistor space localisation is built in.
The method also carries the great practical advantage
of tremendously simplifying the calculation of amplitudes.
It was quickly confirmed that MHV diagrams
gave correct results at tree
level (see \cite{vvkreview} for a review of the field up to
August 2004). The prognosis at one-loop
was initially poor, however, as general arguments
indicated that it would be impossible to
ignore conformal supergravity fields propagating in the loops
\cite{BerkWitt2}.
Separately to this, initial explorations of the
differential equations satisfied by known
one-loop amplitudes appeared to suggest
unexpected complications in their
twistor-space localisation properties \cite{csw2}.

A direct derivation from MHV diagrams
of the one-loop MHV scattering amplitudes in $\cN\!=\!4$ SYM
was presented in \cite{bst}.
The question of the localisation of amplitudes was
then revisited, and the
complications previously found were seen to be due to
the appearance of additional inhomogeneous terms \cite{csw3}
in the differential equations obeyed by one-loop amplitudes.
Taking into account the corrections coming from this,
one finds that indeed the one-loop MHV  amplitudes
in $\cN\!=\!4$ SYM localise on pairs of lines in twistor space
\cite{benabern}, as the direct construction of \cite{bst} suggests.
This encourages one to conjecture that the whole quantum theory of
$\cN\!=\!4$ SYM possesses simple
twistor space localisation properties.
By studying the differential equations satisfied by the
unitarity cuts of amplitudes, the
coefficients of the box functions in the
next to MHV (NMHV) amplitudes
in $\cN\!=\!4$ SYM have recently been shown
to localise on planes in twistor space
\cite{freddy3}. A direct MHV diagram construction
of these amplitudes
has not yet been given however.

The study of the analytic properties of amplitudes, using
the twistor-inspired approach, has since been found useful
in the general analysis of
one-loop amplitudes in $\cN\!=\!4$ Yang-Mills, with a recent
derivation of the $(---++++)$ one-loop NMHV amplitude
\cite{freddy,freddy2}.
This coincides with one case of the
general one-loop seven-gluon NMHV amplitude which
was also found recently \cite{new}
using the cut-constructibility approach.
For $\cN\!=\!1$ theories,
the twistor space structure of  one-loop amplitudes
was studied in \cite{csw2,BBDD,BBDP}
and it was found that the holomorphic anomaly
of unitarity cuts \cite{csw3}
leads to differential equations \cite{BBDD},
in contrast to algebraic equations for $\cN\!=\!4$ \cite{freddy}, obeyed by the
one-loop amplitudes.

MHV diagrams provide a well-defined prescription for the
direct derivation of amplitudes.
It is natural to ask whether the MHV diagram construction
of the one-loop $\cN\!=\!4$  MHV amplitudes
of \cite{bst} can be generalised in other directions --
in particular to theories with less supersymmetry.
This has been confirmed in recent work \cite{bbst,quig},
where the MHV diagram method was shown to correctly reproduce the
known MHV amplitudes for the
$\cN\!=\!1$ chiral multiplet. This result implies
that one-loop MHV amplitudes for all supersymmetric gauge theories
can be derived from MHV diagrams,
and hence
have simple localisation properties in twistor space.

The close relationship between the MHV diagram construction
and unitarity-based methods \cite{Bern:zx},
first seen in \cite{bst},
and the success in applying this method to the
$\cN\!=\!1$ case, encourages the belief that all
cut-constructible amplitudes may be amenable to this new approach.
It is also of great importance
to explore whether MHV diagrams can be
used at loop level in non-supersymmetric theories.%
\footnote{The paper \cite{csw2} discusses
the twistor structure of some non-supersymmetric
one-loop amplitudes and the
possible role of additional vertices in these models.
A recent paper \cite{higgs}
has also developed a generalised
MHV diagram construction
for scattering amplitudes involving
a Higgs boson and gluons.
These amplitudes are described in terms of
a tree-level, non-supersymmetric effective interaction
which arises by integrating out a heavy top quark
in one-loop diagrams.}
These motivations lead one to consider the
one-loop MHV amplitudes in pure Yang-Mills theory.
These amplitudes consist of terms containing cuts, which we
call the cut-constructible part of the amplitude,
plus additional rational terms.
The amplitudes are of great interest, since they are an example of
one-loop $n$-point scattering amplitudes in
QCD, where all external particles and the particle running
in the loop are gluons, and they can be decomposed as
\begin{equation}
\label{decomposition}
{\cal{A}}_{n}^{\textrm{glue}} \ = \ {\cal{A}}_{n}^{{\cal{N}} = 4}
\,  -\, 4\, {\cal{A}}_{n}^{{\cal{N}} = 1 , \, \textrm{chiral}}
\, + \, {\cal{A}}_{n}^{\, \textrm{scalar}} \ .
\end{equation}
The first term describes the contribution of an $\cN\!=\!4$
SYM multiplet to the amplitude.
The second is $-4$ times the contribution of an
$\cN\!=\!1$ chiral multiplet, and the third is a
non-supersymmetric amplitude with only complex scalars
propagating in the loop. In this paper we focus on the
calculation of the final contribution since the other two
are known. A similar supersymmetric decomposition
exists for one-loop gluon scattering amplitudes with massless
quarks or adjoint fermions running in the loop.

The one-loop MHV amplitude in pure Yang-Mills
is known only for two special cases -- when
the two negative-helicity external gluons are adjacent,
the cut-constructible part is known \cite{Bern:1994cg};
and, in the five-gluon case, the full amplitude,
including rational parts, has been calculated
for arbitrary helicity configurations in  \cite{bdk9302280}.
In this paper, we will use MHV diagrams to derive a
compact expression for the cut-constructible part of the
general one-loop MHV multi-gluon amplitude
when there are scalar particles in the loop --
the last term of \eqref{decomposition}.
This generalises the known special cases
with adjacent negative-helicity gluons,
and the five-gluon non-adjacent amplitude.
Moreover, this is the first example of the application
of the MHV diagram approach to non-supersymmetric
loop amplitudes, and provides further evidence
that all cut-constructible (parts of) amplitudes may be derived
using standard MHV diagrams.
Of course, it would be extremely interesting
to extend the MHV diagram method
to obtain the rational pieces.
This might require the construction of suitable
MHV vertices where the
off-shell legs are continued to $4-2\e$ dimensions
or the inclusion of additional
effective vertices as proposed in \cite{csw2}.

The plan for the rest of the paper is as follows.
In Section 2 we present the formal expression for the
one-loop MHV diagrams with a complex scalar running in
the loop, which we use in Section 3 to
rederive the known amplitude when the negative-helicity
gluons are in adjacent positions.
In Section 4 we derive a compact expression for the amplitude
in the case where the negative-helicity gluons are
in arbitrary positions.
Our final result is given by Eq.~\eqref{final2}.
We also briefly comment on the twistor
space structure of our result.
Section 5 is devoted to some consistency checks of our
general amplitude.
Specifically, we show that
it correctly incorporates the adjacent case result
\cite{Bern:1994cg}, also directly reproduced in Section 3,
and  the cut-constructible part of the
five-gluon amplitude computed in
\cite{bdk9302280}.
Finally,  we show that our expression has the expected
infrared singularities.

For further related work on the string theory side, and
on the gauge theory side, see \cite{rsv}--\!\!\cite{gmn}
and \cite{Zhu}--\!\!\cite{ggk} respectively.
\section{The scalar amplitude}

In complete similarity with the ${\cal N}\! = \! 4$ and
${\cal N}\! = \! 1$ cases, see e.g.~\cite{bbst},
we can immediately write down the expression for the scalar
amplitude in terms of MHV vertices as
\beqa
\label{loopint}
\cA_n^{{\rm scalar}}  & = &  \sum_{m_1, m_2, \pm }
\int\! d\cM \
\cA(-l_1^{\mp},m_1,\ldots,i^{-},\ldots,m_2,l_2^{\pm})\
\nonumber \\
&&
\qquad \qquad \qquad
\!\cdot \, \cA(-l_2^{\mp},m_2+1,\ldots,j^{-},
\ldots,m_1-1,l_1^{\pm})
\  ,
\eeqa
where the ranges of summation of $m_1$ and $m_2$ are
\beq
\label{range}
j +1\leq m_1\leq i \ , \qquad
i \leq m_2 \leq j-1 \ .
\eeq
The typical MHV diagram contributing to $\cA_n^{{\rm scalar}}$,
for fixed $m_1$ and $m_2$, is depicted in Figure 1.
\begin{figure} [ht]
\label{scalardiagram}
\vspace{.2in}
\centerline {
\includegraphics[width=4in]{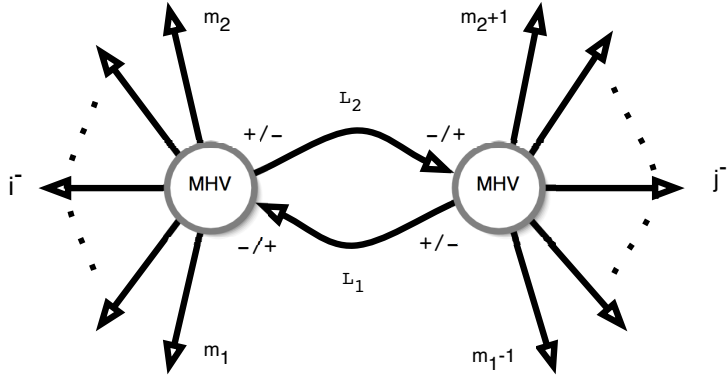}
}
\vspace{.2in}
\caption{\it A one-loop MHV diagram with a complex scalar running in the loop, computed in
Eq.~\eqref{loopint}.
We have indicated the possible helicity assignments
for the scalar particle.}
\end{figure}
The off-shell vertices  $\cA$ in \eqref{loopint}
correspond to having complex scalars running in the loop.
It follows that there are two possible
helicity assignments\footnote{For scalar fields, the  ``helicity''
simply distinguishes
particles from  antiparticles (see, for example, \cite{dixon}).}
for the scalar particles in the loop which have to be summed over.
These two possibilities are denoted
by $\pm$ in \eqref{loopint} and in
the internal lines in Figure 1.
It turns out that each of them
gives rise to the same integrand for \eqref{loopint},
 \beq \label{integrand}
  -i\cA_n^{\rm tree} \,
\cdot \, \frac{\langle m_2\,  m_2 \! + \! 1 \rangle \,
\langle m_1 \! - \! 1
\, m_1 \rangle
\, \langle i\,  l_1 \rangle^2 \, \langle j \,
l_1 \rangle^2 \langle i \, l_2
\rangle^2 \langle
j \, l_2 \rangle^2}{\langle i\,  j\rangle^4 \,
\langle m_1 \, l_1 \rangle \,
 \langle m_1 \! -
\! 1 \, l_1 \rangle \, \langle m_2 \, l_2 \rangle \,
\langle m_2 \! + \! 1
 \, l_2\rangle \,
\langle l_1 \, l_2 \rangle^2} \,\, .
\eeq
A crucial ingredient in \eqref{loopint} is the
integration measure $d \cM$.
This measure was constructed in
\cite{bst},  using the decomposition
$L := l  + z  \eta$ for a non-null four-vector
$L$ in terms of a null vector $l$ and a real parameter $z$.
$\eta$ is a null reference vector,
which disappears in the final result.
We refer the reader to Sections 3 and 4 of \cite{bst}
for the construction of this measure
(also reviewed in Section 3 of
\cite{bbst}), and here we merely quote the result:
\beq
\label{lint}
d\cM  \ = \ \frac{dz}{z} \,
d^{4-2\epsilon}{\rm LIPS}(l_2 , -l_1 ; P_{L;z})\ ,
\eeq
where $L_i := l_i + z_i \eta$, $i=1,2$ and  $z:= z_1 - z_2$.
Thus the integration measure $d \cM$
decomposes into  the product of a
Lorentz-invariant two-particle phase space measure
and a dispersive measure $dz/z$.
The momentum $P_{L;z}$ flowing in the phase space measure is
\beq
P_{L;z} \ := \ P_L \, - \, z \eta
\ .
\eeq
The interpretation of $dz / z $ as a dispersive measure
follows at once when one observes that \cite{bst}
\beq
\label{dz/z}
{dz \over z} \ = \ { d P_{L; z}^2 \over P_{L;z}^2 - P_{L}^2}
\ .
\eeq
In order to calculate \eqref{loopint},
we will first integrate the expression \eqref{integrand}
over the Lorentz invariant phase space (appropriately
regularised to $4-2\e$ dimensions),
and then perform the dispersion integral.
For the sake of clarity,
we will separate the analysis into two parts.
Firstly, we will present the (simpler) calculation
of the amplitude in the case where
the two negative-helicity gluons
are adjacent. This particular amplitude  has already been computed
by Bern, Dixon, Dunbar and Kosower in
\cite{Bern:1994cg} using the cut-constructibility approach;
the result we will derive here
will be in precise agreement with the result in that approach.
Then, in Section 4 we will move on to address the general case,
deriving new results.

\section{The scattering amplitude with
adjacent negative-helicity gluons}

The adjacent case corresponds to choosing
$i = m_1$, $j = m_1-1$ in Figure 1.
Therefore we  now have a single sum over MHV diagrams,
corresponding to the possible choices of $m_2$.
We will also set $i=2$, $j=1$ for the sake of definiteness,
and $m_2 = m$.

After conversion into traces,
the integrand of (\ref{loopint})
takes on the form:
\beqa
\label{reallyfinally}
\frac{
 {\tr}_{+}(\kslash_{1}\, \kslash_{2}\,
\Pslash_{L;z} \, \lslash_{2})
\ {\tr}_{+}(\kslash_{1}\, \kslash_{2}
\, \lslash_{2} \, \Pslash_{L;z})}{2^5 \, (k_1 \cdot k_2)^3 \,
(l_1 \cdot l_2)^2}
\Bigg\{\frac{{\tr}_{+}(\kslash_1 \, \kslash_2\,
\kslash_{m+1} \, \lslash_2)}{(l_2 \cdot m \! + \! 1)}-
\frac{{\tr}_{+}(\kslash_1 \, \kslash_2\,
\kslash_{m} \, \lslash_2)}{(l_2 \cdot m)}\Bigg\} \  ,
\eeqa
where we note that
$(l_1 \cdot l_2) = -P_{L;z}^2/2$ by momentum conservation.

The next step consists of performing
the Passarino-Veltman reduction \cite{pv}
of the Lorentz invariant phase space integral of
\eqref{reallyfinally}.
This requires the calculation of the
three-index tensor integral
\beq
\cI^{\m \n \r} (m, P_{L;z})
\ = \
\int\! d{\rm LIPS}(l_2,-l_1;P_{L;z})
\,
\frac{l_2^{\mu}\, l_2^{\nu} \, l_2^{\r} }{
(l_2\cdot m)}
\ .
\eeq
This calculation is performed in Appendix A.
The result of this procedure gives the following
term at $\cO(\epsilon^0)$, which we will
later integrate with the dispersive measure:
\beqa
\nonumber
\tilde{\cA}_n^{{\rm scalar}} \ &=& \
 \frac{\pi}{3}\,
(-P_{L;z}^2)^{-\e} \,
\frac{\left[{\tr}_{+}(\kslash_{1}\, \kslash_{2}\,
\kslash_{m} \, \Pslash_{L;z})\right]^2}{2^5 \,
(k_1\cdot k_2)^3}\,
\Bigg\{\frac{{\tr}_{+}(\kslash_{1}\, \kslash_{2}\,
\Pslash_{L;z} \, \kslash_{m})}{(m\cdot P_{L;z})^3} \ + \
\frac{2(k_1 \cdot k_2)}{(m\cdot P_{L;z})^2}\Bigg\}
\\
\ &-& \ (m \leftrightarrow m + 1) \ ,
\label{atlast}
\eeqa
and we have dropped a factor of $4\pi \l \cA^{\rm tree}$
on the right hand side of
\eqref{atlast}, where $\lambda$ is defined in \eqref{ubi}.
We can reinstate this factor at the end of the calculation.
We also notice that  \eqref{atlast} is a finite expression,
i.e.~it is free of infrared poles.

An important remark is in order here.
On general grounds, the result of a phase space
integral in, say, the $P^2$-channel, is of the form
\beq
\label{phasespaceint}
\cI( \e) \ = \ (-P^2)^{-\e} \cdot f (\e)
\ ,
\eeq
where
\beq
\label{effe}
f ( \e) \ = \ {f_{-1} \over  \e} \, + \,
f_0 \, + \, f_1 \e \, + \, \cdots
\ ,
\eeq
and $f_i$ are rational coefficients.
In the case at hand, infrared poles generated
by the phase space integrals cancel completely,
so that we can in practice replace
\eqref{effe} by
$f(\e)  \to  f_0 \, + \, f_1 \e
\, + \, \cdots $.
The amplitude $\cA$ is then obtained
by performing a dispersion integral, which converts
\eqref{phasespaceint} into an expression of the form
\beq
\label{disppint}
\cA( \e) \ = \ {(-P^2)^{-\e} \over \e}  \cdot g (\e)
\ = \ {g_0 \over \e} \, - \, g_0 \log ( -P^2)
\, + \, g_1 \, + \, \cO( \e)
\ ,
\eeq
where $g (\e) = g_0 + g_1 \e + \cdots$, and the coefficients
$g_i$ are rational functions, i.e.~they are free of cuts.
Importantly, errors can be generated in the evaluation
of phase space integrals if one contracts
$(4-2 \e)$-dimensional vectors with ordinary four-vectors.
This does not affect
the evaluation of the coefficient $g_0:=g(\e=0)$,
and hence the part of the amplitude containing cuts
is reliably computed;
but the the coefficients $g_i$ for
$i \geq 1$, in particular $g_1$, are in general affected. This implies that
{\it rational} contributions to the scattering amplitude
 cannot be detected \cite{Bern:1994cg} in this construction.
A notable exception to this is provided by the phase space integrals
which appear in supersymmetric theories. These are
``four-dimensional cut-constructible'' \cite{Bern:1994cg},
in the sense that the rational parts are unambiguously
linked to the discontinuities across cuts,
and can therefore be uniquely determined.\footnote{For more details about cut-constructibility,
see the detailed analysis  in Sections 3-5 of \cite{Bern:1994cg}.}
This occurs, for example,
in the calculation of the $\cN \! = \! 4$
MHV amplitudes at one loop performed in \cite{bst}.
In the present case, however,
the relevant phase space integrals violate
the cut-constructibility criteria given in
\cite{Bern:1994cg}%
\footnote{An example of an
integral violating the power-counting criterion of \cite{Bern:1994cg} is provided
by \eqref{imunu}.}, since we encounter tensor triangles with up to
three loop momenta in the numerator.
Hence, we will be able to compute the part
of the amplitude containing cuts, but not the rational terms.
In practice, this means that we will compute
all phase space integrals up to $\cO (\e^0 )$,
and discard $\cO (\e)$ contributions, which would generate rational terms
that cannot be determined correctly.

After this digression,
we now move on to the dispersion integration.
In the center of mass frame, where
$P_{L;z}\! :=\! P_{L;z}(1,\textbf{0})$,
all the dependence on $P_{L;z}$ in \eqref{atlast}
cancels out, as there are equal powers of $P_{L;z}$
in the numerator as in the denominator
of any term. As a consequence, the dependence on the
arbitrary reference vector $\eta$ disappears (see \cite{quig} for
the application of this argument to the $\cN=1$ case).
Using \eqref{dz/z} in order to re-express $dz/z$ in terms of the
relevant dispersive measure, we see that
we are left with dispersion integrals
of the form
\beqa
\label{disperse}
I(P_L^2) \ := \ \int\! \frac{ds'}{s'-P_L^2}\,
(s')^{-\epsilon} \ = \
\frac{1}{\epsilon}[\pi\epsilon
\csc(\pi\epsilon)]\, (-P_L^2)^{-\epsilon}
\ .
\eeqa
Taking this into account,
the dispersion integral of  \eqref{atlast}
then gives
\beqa
\nonumber
\!\!\!\tilde{\cA}_n^{{\rm scalar}} \ &=& \
\big[ {\pi \e \csc ( \pi \e) }\big]
\,
\frac{\pi}{3}\,
{(-P_{L}^2)^{-\e}\over \e} \,
\frac{\left[{\tr}_{+}(\kslash_{1}\, \kslash_{2}\,
\kslash_{m} \, \Pslash_{L})\right]^2}{2^5 \, (k_1\cdot k_2)^3}\,
\Bigg[
\frac{{\tr}_{+}(\kslash_{1}\, \kslash_{2}\,
\Pslash_{L} \, \kslash_{m})}{(m\cdot P_{L})^3} \ + \
\frac{2(k_1 \cdot k_2)}{(m\cdot P_{L})^2}\Bigg]
\\
\ &-& \ (m \leftrightarrow m + 1) \ .
\label{atlast2}
\eeqa
The momentum flow can be conveniently represented as in
Figure 2, where we define
\beq
P \ := \ q_{2,m-1} \ , \qquad Q \ := \ q_{m+1, 1}
\ = \ -\, q_{2,m}
\ ,
\eeq
and $q_{p_1, p_2} := \sum_{l=p_1}^{p_2} k_l$.
We also have $P_L := q_{2,m} = -Q$.

Now we wish to combine the terms in the first line of
\eqref{atlast2} with those
in the second line.
Since \eqref{atlast2} is summed over $m$,
we simply shift  $m+1 \to m$ in the terms of the
second line.
Let us now focus our attention on
the second  term in \eqref{atlast}
(similar manipulations
will be applied to the first term).
Writing the $m \lrar m+1$ term explicitly,
we obtain a contribution proportional to
\beq
\label{typical}
(-P_L^2 )^{-\epsilon}
\bigg[
\frac{\left[{\tr}_{+}(\kslash_{1}\, \kslash_{2}\,
\kslash_{m} \, \Pslash_L)\right]^2}{(m\cdot P_L )^2} \ - \
\frac{\left[
{\tr}_{+}(\kslash_{1}\, \kslash_{2}\,
\kslash_{m+1} \, \Pslash_{L})\right]^2}{((m+1)\cdot P_{L})^2}
\biggr]
\ .
\eeq
By shifting  $m+1 \to m$ in the second term
of \eqref{typical},
we convert its $P_L$  to $P_L \to q_{2,m-1} = P$
(whereas, in the non-shifted term, $P_L = -Q$).
The expression \eqref{typical} then reads
\beq
\frac{\left[{\tr}_{+}(\kslash_{1}\, \kslash_{2}\,
\kslash_{m} \, \Qslash)\right]^2}{(m\cdot Q )^2}
\,
\Big[ (-Q^2)^{-\e} - (-P^2)^{-\e}\Big]
\ ,
\eeq
where we used ${\tr}_{+}(\kslash_{1}\, \kslash_{2}\,
\kslash_{m} \Qslash)  = - {\tr}_{+}(\kslash_{1}\, \kslash_{2}\,
\kslash_{m}\Pslash)$ and
$Q\cdot m= - P \cdot m$.
Notice also that $m\cdot Q  = - (1/2) (Q^2 - P^2)$.

\begin{figure} [ht]
\label{adjacent}
\vspace{.2in}
\centerline {
\includegraphics[width=4in]{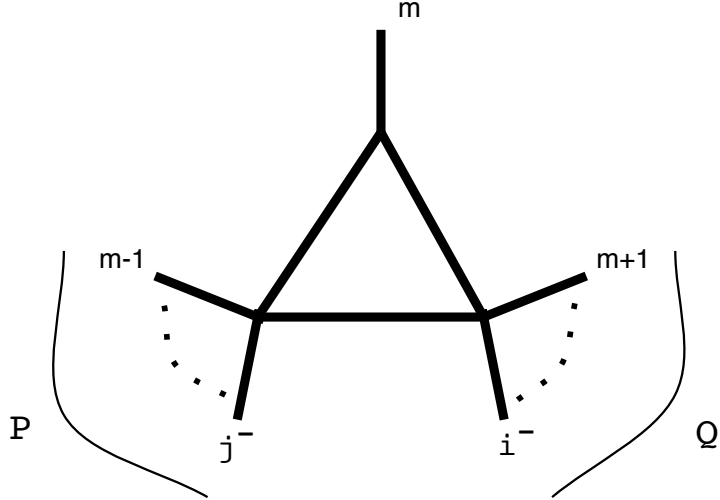}
}
\vspace{.2in}
\caption{\it A triangle function contributing to the amplitude
in the case of adjacent negative helicity gluons.
Here we have defined $P :=  q_{j,m-1}$,
$Q  :=  q_{m+1, i}=  - q_{j,m}$
(in the text we set $i=1$, $j=2$ for definiteness).
}
\end{figure}
Next we re-instate the antisymmetry of the amplitudes under
the exchange of the indices $1 \lrar 2$ (which is manifest from
equation \eqref{integrand}).
Doing this we get
\beqa
\big[
{\tr}_{+}(\kslash_{1}\, \kslash_{2}\, \kslash_{m} \Qslash)
\big]^2 &\longrightarrow&
{1\over 2}
\Big[
\big({\tr}_{+}(\kslash_{1}\, \kslash_{2}\, \kslash_{m} \, \Qslash)
\big)^2 \, - \,
\big({\tr}_{+}(\kslash_{1}\, \kslash_{2}\,  \Qslash \, \kslash_{m})
\big)^2
\Big]
\\ [6pt]\nonumber
&=& 2 (k_1 \cdot k_2) (m\cdot Q)
\Big[
{\tr}_{+}(\kslash_{1}\, \kslash_{2}\, \kslash_{m} \, \Qslash)
\, - \,
{\tr}_{+}(\kslash_{1}\, \kslash_{2}\,  \Qslash \, \kslash_{m})
\Big]
\ .
\eeqa
Following similar steps for the first term in
\eqref{atlast2}, we arrive at the following expression
for the amplitude before taking the $\e \to 0$ limit:
\beq
\cA_\e \ = \ \cA_{1, \e} \, + \, \cA_{2, \e}
\ ,
\eeq
where
\beqa
\nonumber
\cA_{1, \e} & =&  - {\cA^{\rm tree} \over  t_1^{[2]}} \cdot
{\pi \over 6}
\cdot
\Bigl[
{\tr}_{+}(\kslash_{1}\, \kslash_{2}\, \kslash_{m} \,
\qslash_{m,1})
\, - \,
{\tr}_{+}(\kslash_{1}\, \kslash_{2}\,
\qslash_{m,1}\, \kslash_{m} )
\Bigr] \, T_{\e} (m, q_{2, m-1} , q_{2 , m})
\ ,
\\ \nonumber \cr
\label{adue}
\cA_{2, \e}  &=&
- {\cA^{\rm tree} \over (t_1^{[2]})^3} \cdot {\pi \over 3}
\cdot
\Bigl[
\big[
{\tr}_{+}(\kslash_{1}\, \kslash_{2}\, \kslash_{m} \,
\qslash_{m,1})\bigl]^2
{\tr}_{+}(\kslash_{1}\, \kslash_{2}\, \qslash_{m,1}
\,\kslash_{m}  )
\, - \,
\\ [6pt]  
&-&{\tr}_{+}(\kslash_{1}\, \kslash_{2}\, \kslash_{m} \,
\qslash_{m,1})
\big[ {\tr}_{+}(\kslash_{1}\, \kslash_{2}\,
\qslash_{m,1}\, \kslash_{m} )\big]^2
\Bigr] \,
T_{\e}^{(3)} (m, q_{2, m-1} , q_{2 , m})
\ ,
\eeqa
and $t_{1}^{[2]}$ follows from the definitions below
equation (\ref{finbox}). In order to write \eqref{adue}
in a compact from, we have introduced
$\epsilon$-dependent triangle functions
\cite{bbst}
\beq
\label{epstriangle}
T_{\epsilon}^{(r)} (p, P, Q) \ := \
{1\over \epsilon}
{ (-P^2)^{-\epsilon} - (-Q^2)^{-\epsilon} \over (Q^2 - P^2)^r}
\ ,
\eeq
where  $p+P+Q=0$, and $r$ is a positive integer.%
\footnote{For $r=1$ we will omit the superscript $(1)$
in $T^{(1)}$. }

We can now take the $\e \to 0$ limit.
As long as $P^2$ and $Q^2$ are non-vanishing, one has
\beq
\label{TT1}
\lim_{\epsilon \to 0} T_{\epsilon}^{(r)} (p, P, Q) \ = \
T^{(r)} (p, P, Q) \ , \qquad
P^2 \neq 0 \, , \, Q^2 \neq 0
\ ,
\eeq
where the $\epsilon$-independent triangle functions are defined by
\beq
\label{trianr}
T^{(r)} (p, P, Q) \ := \
{ \log (Q^2 / P^2) \over ( Q^2 - P^2)^r}
\ .
\eeq
If either of the invariants vanishes,
the limit of the $\e$-dependent triangle
gives rise to an infrared-divergent term
(which we call a ``degenerate'' triangle -
this is one with two massless legs).
For example, if
$Q^2=0$, one has
\beq
\label{TT2}
\left. T_{\epsilon} (p, P, Q)
\right|_{Q^2=0}  \ \longrightarrow \
- {1\over \epsilon} \, {(-P^2)^{-\epsilon }\over P^2}\, ,
\qquad {\epsilon \to 0}
\ .
\eeq
The two possible configurations which give rise to
infrared divergent contributions
correspond to the following two possibilities:
\begin{itemize}
\item[{\bf a.}]
$q_{2, m-1} = k_2 $ (hence $q_{2, m-1}^2 = 0$).
In this case we also have
$q_{2,m}^2 = t_{2}^{[2]}$.
\item[{\bf b.}]
$- q_{2, m} = k_1$ (hence $q_{2, m}^2 = 0$).
Therefore  $q_{2, m-1}^2 = t_{n}^{[2]}$.
\end{itemize}
We notice that infrared poles will appear only in terms
corresponding to the triangle function  $T$.
Indeed, whenever one of the
kinematical invariants contained in $T^{(3)}$ vanishes,
the combination of traces multiplying this function in
\eqref{adue} vanishes as well.

In conclusion, we arrive at the following result,
where we have explicitly separated  out the infrared-divergent
terms:%
\footnote{A factor of
$-4 \pi \l$ will be understood on the right hand sides
of Eqs.~\eqref{sz}, \eqref{endresult}, \eqref{chiral},
where  $\l$ is defined in \eqref{ubi}.}
\beq
\label{sz}
\cA_{n}^{\rm scalar} \ = \
\cA_{\rm poles} \, + \,
\cA_1 \, + \, \cA_2
 \ ,
\eeq
where
\beqa
\cA_{\rm poles} & = &
{1\over 6} \cA^{\rm tree} {1 \over \e }
\Bigl[ (-t_2^{[2]})^{-\e} \, + \, (-t_n^{[2]})^{-\e}
\Bigr]
\ ,
\\ [6pt] \nonumber
\cA_{1} & = &
 {1\over 6} \cA^{\rm tree}
{1\over t_1^{[2]}} \,
\sum_{m=4}^{n-1} \Bigl[
{\tr}_{+}(\kslash_{1}\, \kslash_{2}\, \kslash_{m} \,
\qslash_{m,1})
\, - \,
{\tr}_{+}(\kslash_{1}\, \kslash_{2}\,
\qslash_{m,1}\, \kslash_{m} )
\Bigr] \, T  (m, q_{2, m-1} , q_{2 , m})
\ ,
\\ [6pt] \nonumber
\cA_2  &=&
{1\over 3}
\cA^{\rm tree}
{1 \over (t_1^{[2]})^3}
\sum_{m=4}^{n-1}
\Bigl[
\big[ {\tr}_{+}(\kslash_{1}\, \kslash_{2}\, \kslash_{m} \,
\qslash_{m,1})\bigl]^2
{\tr}_{+}(\kslash_{1}\, \kslash_{2}\, \qslash_{m,1}
\,\kslash_{m}  )
\\ [6pt]\nonumber \cr
&-&{\tr}_{+}(\kslash_{1}\, \kslash_{2}\, \kslash_{m} \,
\qslash_{m,1})
\big[ {\tr}_{+}(\kslash_{1}\, \kslash_{2}\,
\qslash_{m,1}\, \kslash_{m} )\big]^2
\Bigr] \,
T^{(3)} (m, q_{2, m-1} , q_{2 , m})
\ .
\eeqa
More compactly, we can recognise that
$\cA_{\rm poles}$ and $\cA_{1}$  reconstruct
the contribution of an $\cN\!=\!1$
chiral supermultiplet,  and rewrite
\eqref{sz} as
\beqa
\label{endresult}
\cA_n^{\textrm{scalar}} \ = \
\frac{1}{3}\cA_{12}^{\cN = 1,\, \textrm{chiral}}
\ + \ \frac{1}{3}\cA_{12}^{\textrm{tree}}
\frac{1}{(t_1^{[2]})^3}\sum_{m=4}^{n-1}
\cB_{12}^{m}\,
T^{(3)}  (m, q_{2, m-1} , q_{2 , m})
\ ,
\eeqa
where
\beqa
\label{scalar2}
\cB_{12}^{m} &=&
 \bigl[{\tr}_{+}(\kslash_{1}\, \kslash_{2}\,
\kslash_{m} \, \qslash_{m,1})\bigr]^2{\tr}_{+}(\kslash_{1}\,
\kslash_{2}\, \qslash_{m,1} \, \kslash_{m})
\\ \nonumber \cr
&-&
\bigl[{\tr}_{+}(\kslash_{1}\, \kslash_{2}\,
\qslash_{m,1} \, \kslash_{m})\bigr]^2{\tr}_{+}(\kslash_{1}\,
\kslash_{2}\,
\kslash_{m} \, \qslash_{m,1})
\ .
\eeqa
and
\beq
\label{chiral}
\cA_{12}^{\cN = 1,\, \textrm{chiral}} \ = \
\frac{1}{2}\cA_{12}^{\textrm{tree}}\frac{1}{t_1^{[2]}}\
\sum_{m=3}^{n}\Big[{\tr}_{+}(\kslash_{1}\, \kslash_{2}\,
\kslash_{m} \, \qslash_{m,1}) \, - \, {\tr}_{+}(\kslash_{1}\,
\kslash_{2}\,
\qslash_{m,1} \, \kslash_{m})\Big]\
T (m, q_{2, m-1} , q_{2 , m})
\ .
\eeq
This is our result for the cut-constructible part of
the $n$-gluon MHV scattering amplitude with adjacent
negative-helicity gluons in positions 1 and 2.
This expression was first derived
by Bern, Dixon, Dunbar and Kosower in \cite{Bern:1994cg},
and our result agrees precisely with this.
A remark is in order here.
In \cite{Bern:1994cg}, the final result is
expressed in terms of a function
\beq
L_2 (x) \ := \ {\log x - (x - 1/x)/2 \over (1-x)^3}
\ ,
\eeq
which contains a rational part $ -  (x - 1/x) / 2(1-x)^3$.
This rational part removes a spurious
third order pole from the amplitude, but
with our approach we did not expect to detect rational
terms in the scattering amplitude, and indeed
we do not find  such terms\footnote{In our notation $L_2$
corresponds to $T^{(3)}$, which, however, lacks a rational term.}.
Furthermore, we do not find the other rational terms
which are known to be present in the one-loop
scattering amplitude \cite{bdk9302280}.


\section{The scattering amplitude in the general case}


The situation where the negative-helicity gluons are not
adjacent is technically more challenging.
Our starting point will be
\eqref{integrand}, to which we will apply the Schouten identity
(see Appendix D for a collection
of spinor identities used in this paper).
Eq.~\eqref{integrand}
can then be written as a sum of four terms:%
\footnote{We drop the factor of
$-i\cA_n^{\rm tree}$ from now on and
reinstate it at the end of the calculation.}
\beqa
\label{4integrand} {\cal C}(m_1, m_2  +  1) \ - \
{\cal C}(m_1,m_2)
\ - \  {\cal C}(m_1-1, m_2 +  1)\  +\ {\cal C}(m_1  - 1, m_2)
\ ,
\eeqa
where
\beqa
\label{basicC}
{\cal C}(a,b) \ := \
\frac{\langle i \, l_1 \rangle \,
\langle j \, l_1 \rangle^2 \,
\langle i \, l_2 \rangle^2 \,
\langle j \, l_2 \rangle}{\langle i \, j \rangle^4 \,
\langle l_1 \, l_2 \rangle^2}
\, \cdot \, \frac{\langle i \, a \rangle \, \langle j \, b \rangle}
{\langle l_1 \, a \rangle \, \langle l_2 \, b \rangle}\ .
\eeqa
The calculation of the phase space integral of this expression is
discussed in Appendix B.
The result is
\begin{eqnarray}
 \label{pspaceresult}
\nonumber
&&  \!\!\!\!\!\!\!\!\!\!\!\!\!\!\!\!\!\!\!\!\!\!\!\!
\int \!d^{4-2\epsilon}{\rm LIPS}(l_2 , -l_1 ; P_{L;z})\ \
{\cal C}(a,b)   \\ [6pt]
\nonumber \cr
&& \quad = \ \frac{1}{3} \
\frac{\ijba}{(a\cdot b)}\ \Bigg[ \ijPLza^2\bigg[
\frac{\ijaPLz}{(P_{L;z}\cdot a)^3}
+ \frac{2(i\cdot j)}{(P_{L;z}\cdot a)^2} \bigg]  -
( a \leftrightarrow b)   \Bigg]\\ [6pt]\nonumber \cr
   &&  \quad\qquad + \
\frac{1}{2}\
\frac{\ijba\ijab}{(a\cdot b)^2}
      \  \Bigg[ \frac{\ijPLza^2}{(P_{L;z}\cdot a)^2} +
( a \leftrightarrow b) \Bigg]
\\ [6pt]\nonumber  \cr
&&   \qquad \qquad - \ \frac{\ijab\ijba}{(a \cdot b)^3}\
\Bigg[  \frac{\ijab\ijPLza}{(P_{L;z}\cdot a)}
               + ( a \leftrightarrow b)  \Bigg]
\\ [6pt] \cr
    && \qquad\qquad\qquad +\
%
%
\frac{\ijab^2\ijba^2}{(a\cdot b)^4} \
    \log\bigg( 1 - \frac{(a\cdot b)}{N}P_{L;z}^2 \bigg)
\ ,
\end{eqnarray}
where $N := N(P) := (a\cdot b)P^2 - 2(P\cdot a)(P \cdot b)$,
and we have suppressed a factor of
$-4 \pi \l (-P_L^2)^{-\e}
\cdot [2^8(i\cdot j)^4]^{-1}$
on the right hand side of \eqref{pspaceresult}, where
$\l$ is defined in \eqref{ubi}.
We notice that  \eqref{pspaceresult} is symmetric under the
simultaneous exchange of $ i$ with $j$ and $a$ with $b$.
This symmetry is manifest in the coefficient multiplying the
logarithm -- the last term in \eqref{pspaceresult};
for the remaining terms, nontrivial gamma matrix identities
are required.
For instance, consider the terms in the second line of
 \eqref{pspaceresult}. These terms are present in
the adjacent gluon case \eqref{atlast}, and it is therefore
natural to expect that the trace structure of this term
is separately invariant when $i \lrar j$ and $a \lrar b$.
Indeed this is the case, thanks to the identity
\beqa
32 (i \cdot j )^3 & = & \ijPLza^2\bigg[
\frac{\ijaPLz}{(P_{L;z}\cdot a)^3}
\, + \, \frac{2(i\cdot j)}{(P_{L;z}\cdot a)^2} \bigg]
\\ \nonumber
& + &
\ijaPLz^2\bigg[
\frac{\ijPLza}{(P_{L;z}\cdot a)^3}
\, + \, \frac{2(i\cdot j)}{(P_{L;z}\cdot a)^2} \bigg]
 \ .
\eeqa
Similar identities show that the third and fourth line of
\eqref{pspaceresult} are invariant under
the simultaneous exchange $i \lrar j$ and $a \lrar b$.

The next step is to perform the dispersion integral of
\eqref{pspaceresult}, i.e.~the integral
over the variable $z$.
This appears in the terms involving
$P_{L;z}$ in \eqref{pspaceresult},
and in an overall factor $(P_{L;z}^2)^{-\epsilon}$
arising from the dimensionally regulated measure.

The integral over the term involving the
logarithm has been evaluated in \cite{bst}, with the
result
\beqa
\nonumber
\label{dilogint}
\int \frac{dz}{z}  (P_{L;z}^2)^{-\epsilon}
\log\left( 1 - \frac{(a\cdot b)}{N}P_{L;z}^2 \right)
 &=&
\int \frac{dP_{L;z}^2 }{P_{L;z}^2 - P_{L}^2 }
(P_{L;z}^2)^{-\epsilon}
\log\bigg( 1 - \frac{(a\cdot b)}{N}P_{L;z}^2 \bigg)
\\ \cr
&=&
\Li\left( 1- \frac{(a\cdot b)}{N(P)} P_L^2 \right)
\ + \ \cO ( \e )
\ .
\eeqa
Notice that these terms were not present in the
adjacent negative-gluon case considered in Section 3.

Next we move  on to the remaining terms
in \eqref{pspaceresult}.
Inspecting their $z$-dependence,
we see that, in complete similarity
with the adjacent case of Section 3,
in each term there are the same powers of $P_{L;z}$
in the numerator as in the denominator.
Hence, in the centre of mass frame in which
$P_{L;z}\! :=\! P_{L;z} (1,\textbf{0})$, one finds that
$P_{L;z}$ cancels completely.
Note that this also immediately resolves
the question of gauge invariance for these
terms -- this occurs only through the $\eta$ dependence in
$P_{L;z} = P_L -z\eta$.
Furthermore, the  box functions coming from \eqref{dilogint}
are separately gauge invariant \cite{bst}.
The conclusion is that our expression for the amplitude below,
built from sums over MHV diagrams of the
dispersion integral
of \eqref{pspaceresult}, will be gauge invariant.
Moreover, apart from \eqref{dilogint},
the only other dispersion integral we will need
is that computed in \eqref{disperse}.

It follows from this discussion that the result
of the dispersion integral of
\eqref{pspaceresult} is (suppressing a factor of
$-4 \pi \l (-P_L^2)^{-\e}
\cdot [2^8(i\cdot j)^4]^{-1} \cdot
[\pi\epsilon \csc(\pi\epsilon)]$):
\begin{eqnarray}
 \label{dispspaceresult} \nonumber
&&  \!\!\!\!\!\!\!\!
\int \frac{dz}{z}\, \int \!
d^{4-2\epsilon}{\rm LIPS}(l_2 , -l_1 ; P_{L;z})\ \
{\cal C}(a,b)
\\ [6pt] \nonumber \cr
&& \quad  = \
 \ \frac{1}{\epsilon}\ {(-P_L^2)^{-\epsilon}}
\ \Bigg\{
    \ \frac{1}{3} \ \frac{\ijba}{(a\cdot b)}\
\Bigg[ \ijPLa^2\bigg[  \frac{\ijaPL}{(P_L\cdot a)^3}
+ \frac{2(i\cdot j)}{(P_L\cdot a)^2} \bigg]  -
( a \leftrightarrow b)   \Bigg]
\\ [6pt]\nonumber \cr
   &&  \quad\qquad + \ \frac{1}{2}\ \frac{\ijba\ijab}{(a\cdot b)^2}
      \  \Bigg[ \frac{\ijPLa^2}{(P_L\cdot a)^2} +
( a \leftrightarrow b) \Bigg]
\\ [6pt]\nonumber  \cr
&&   \qquad \qquad - \ \frac{\ijab\ijba}{(a\cdot b)^3}\
\Bigg[  \frac{\ijab\ijPLa}{(P_L \cdot a)}
               + ( a \leftrightarrow b)  \Bigg]  \  \Bigg\}
\\ [6pt]\cr
    && \qquad\qquad\qquad +\
\frac{\ijab^2\ijba^2}{(a\cdot b)^4} \
    \Li\left( 1 - \frac{(a\cdot b)}{N(P_L)}P_L^2 \right)\ .
\end{eqnarray}

Now, due to the four terms in \eqref{4integrand},
the sum over MHV diagrams will include
a signed sum over four expressions like
\eqref{dispspaceresult}.
Let us begin by considering
the last line of \eqref{dispspaceresult}.
This is a term familiar from \cite{bst} and
\cite{bbst}, corresponding to
one of the four dilogarithms in
the novel expression found in
\cite{bst} for the finite part $B$
of a  scalar box function,
\beqa
\label{boxf}
\nonumber
B(s,t,P^2,Q^2) & = &
\Li\left(1-\frac{(a\cdot b)}{N(P)}P^2\right) \, + \,
\Li\left(1-\frac{(a\cdot b)}{N(P)}Q^2\right)
\\  \cr
&-&
\Li\left(1-\frac{(a\cdot b)}{N(P)}s\right)
\, - \,
\Li\left(1-\frac{(a\cdot b)}{N(P)}t\right)
\ ,
\eeqa
with $s:= (P + a)^2$,  $t:= (P + b)^2$, and
$P + Q + a + b = 0$.
By taking into account the four terms in
\eqref{4integrand} and summing over MHV  diagrams
as specified in \eqref{loopint} and \eqref{range},
one sees that each of the four terms in any finite
box function $B$ appears exactly once,
in complete similarity with \cite{bst}
and \cite{bbst}, so that the final contribution of this term
will be%
\footnote{We multiply our final results by a factor of $2$,
which takes into account the two possible helicity
assignments for the scalars in the loop.}
\beq
\label{finbox}
\sum_{m_1=j+1}^{i-1}\sum_{m_2=i+1}^{j-1}
{1\over 2} \, \big[ b_{m_1  m_2}^{ij} \big]^2
\,
B( q_{m_1, m_2 -1}^2 , q_{m_1+1,  m_2}^2,
q_{m_1+1 , m_2 -1}^2, q_{m_2+1, m_1 -1}^2)
\ ,
\eeq
where $t_i^{[k]} := (p_i + p_{i+1} + \cdots + p_{i+k-1})^2$
for $k\geq 0$, and $t_i^{[k]} =t_i^{[n-k]}$ for $k <0$.
In writing \eqref{finbox},
we have taken into account
that the dilogarithm in \eqref{dispspaceresult}
is multiplied by a coefficient proportional to
the square of $b_{m_1 m_2}^{i j}$, where
\beq
\label{bdef}
b_{m_1 m_2}^{i j} \ :=  \
-2 \, \frac{
\tr_{+} \left( \kslash_{i} \kslash_{j} \kslash_{m_1}
\kslash_{m_2} \right)
\tr_{+} \left( \kslash_{i} \kslash_{j} \kslash_{m_2}
\kslash_{m_1} \right) }
{ [ (k_i + k_j)^{2}]^2
\,
[( k_{m_1} + k_{m_2})^2]^2}
\ .
\eeq
We notice that $b_{m_1 m_2}^{i j}$ is the coefficient
of the box functions in the one-loop
$\cN \! = \! 1$ MHV amplitude, originally calculated by
Bern, Dixon, Dunbar and Kosower in \cite{Bern:1994cg}, and
derived in \cite{bbst, quig}
using the MHV diagram approach
for loops proposed in \cite{bst}.
\begin{figure} [ht]
\label{scalarbox}
\vspace{.2in}
\centerline {
\includegraphics[width=4in]{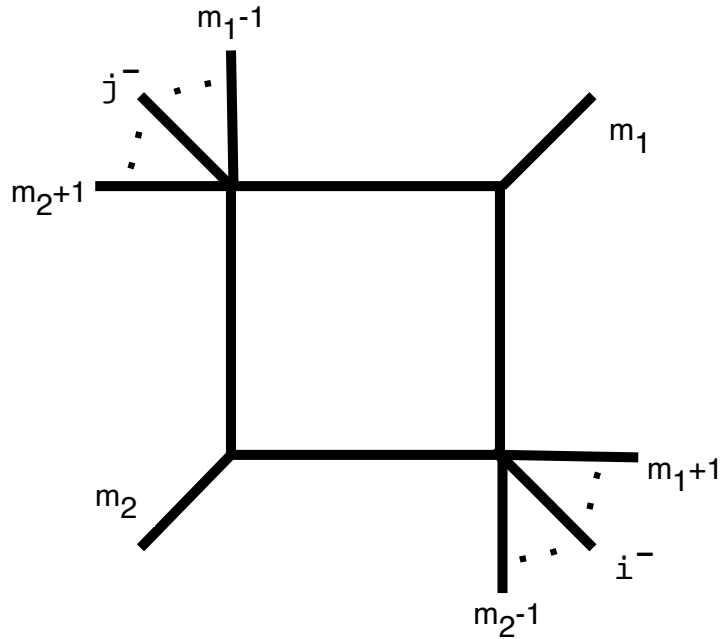}
}
\vspace{.2in}
\caption{\it A box function contributing to the amplitude
in the general case.
The negative-helicity gluons, $i$ and $j$,
cannot be in adjacent positions, as the figure
shows.}
\end{figure}
Furthermore, we observe that $b_{m_1 m_2}^{i j}$ is holomorphic
in the spinor variables, and as such has simple localisation
properties in twistor space.
Indeed, from \eqref{bdef} it follows that
\beq
\label{bdefholo}
b_{m_1 m_2}^{i j} \ =  \
2\,
{ \lan i  \, m_1 \ran \, \lan i  \, m_2 \ran \,
\lan j  \, m_1 \ran \, \lan j  \, m_2 \ran
\over
\lan i  \, j \ran^2  \, \lan m_1  \, m_2 \ran^2
}
\ .
\eeq
Summing over the four terms for the remainder of
\eqref{dispspaceresult} can be done in complete
similarity with Section 4 of \cite{bbst}.%
\footnote{In Section 3 we have illustrated in detail
how this sum is performed for the
simpler case of adjacent negative-helicity gluons.}
We will skip the details of this derivation,
and  will now present our result.

In order to do this, we find it convenient to define
the following expressions:
\beqa
A^{ij}_{m_1 m_2} & := &
{(i\, j\, m_2+1\,  m_1) \over ((m_2 +1) \cdot m_1)} \, - \,
{(i\, j\, m_2\,  m_1) \over (m_2  \cdot m_1)}
\\ [4pt]\nonumber
& = & -2\, [i \, j] \, \lan m_1 \, i \ran
\lan m_1 \, j \ran \,
{\lan m_2 \, m_2 + 1 \ran \over
\lan m_2+1  \, m_1  \ran
\,
\lan m_1 \, m_2  \ran}
\ ,
\\ [4pt]\cr
S^{ij}_{m_1 m_2} & := &
{(i\, j\, m_1\,  m_2 +1)(i\, j\, m_2+1\,  m_1)
\over ((m_2 +1) \cdot m_1)^2} \, - \,
{(i\, j\, m_1\,  m_2) (i\, j\, m_2\,  m_1)
\over (m_2  \cdot m_1)^2}
\ ,
\\[4pt]
I^{ij}_{m_1 m_2} & := &
{(i\, j\, m_1\,  m_2 +1)^2 (i\, j\, m_2+1\,  m_1)
\over ((m_2 +1) \cdot m_1)^3} \, - \,
{(i\, j\, m_1\,  m_2)^2 (i\, j\, m_2\,  m_1)
\over (m_2  \cdot m_1)^3}
\ ,
\eeqa
where for notational simplicity we set
$(a_1\, a_2\, a_3\, a_4) :=
{\tr}_{+} (\aslash_1 \aslash_2 \aslash_3  \aslash_4)$ in the above.
We also note the symmetry properties
\beq
A^{ji}_{m_1 m_2} \, = \, - A^{ij}_{m_1 m_2}
\ ,
\qquad
S^{ji}_{m_1 m_2} \, = \,  S^{ij}_{m_1 m_2}
\ .
\eeq
The momentum flow is best described using the triangle
diagram in Figure 4, where we use the following definitions:
\beqa
P & := & q_{m_2 + 1 , m_1 -1} \ = \ - q_{m_1 , m_2}
\ ,
\\ \nonumber
Q &: = & q_{m_1 + 1 ,  m_2}
\ .
\eeqa
The triangle in Figure 5 also appears in the calculation,
and can be  converted into a triangle as in  Figure 4 --
but  with $i$ and $j$ swapped --
if one shifts $m_1-1 \to m_1$,
and then swaps  $m_1 \lrar m_2$.

We then introduce the coefficients
\beqa
\label{alfredina}
\cA^{ij}_{m_1 m_2} &:= &
2^{-8}
(i\cdot j)^{-4}
\, A^{ij}_{m_1 m_2}
\, \Big[
(i\, j\, m_1\,  Q)^2 (i\, j\,  Q \, m_1) \, - \,
(i\, j\, m_1\,  Q) (i\, j\,  Q \, m_1)^2\Big]
,
\\ \cr
\tilde{\cA}^{ij}_{m_1 m_2} &:= &
2^{-8} (i\cdot j)^{-4}  \, A^{ij}_{m_1 m_2}
\, \Big[
(i\, j\, m_1\,  Q)^2  \, - \,  (i\, j\,  Q \, m_1)^2\Big]
\, ,
\\ \cr
\cS^{ij}_{m_1 m_2} & = &
2^{-8} {(i\cdot j)^{-4}} \, S^{ij}_{m_1 m_2}
\, \Big[
(i\, j\, m_1\,  Q)^2 \, + \,
(i\, j\,  Q \, m_1)^2\Big]
\ ,
\\ \cr
\cI^{ij}_{m_1 m_2} &:= &
2^{-8} (i\cdot j)^{-4}  \, \Big[
I^{ij}_{m_1 m_2}
\, (i\, j\, Q \,  m_1)  \, + \,
I^{ji}_{m_1 m_2}(i\, j\,  m_1 \, Q )\Big]
\label{giuseppina}
\ .
\eeqa
We will also make use of the
$\epsilon$-dependent triangle functions introduced
in \eqref{epstriangle}, whose $\e \to 0$ limits have been
considered in \eqref{TT1}--\eqref{TT2}.
This is in order to write a compact expression
which incorporates also the infrared-divergent terms.%
\footnote{The infrared-divergent terms will be described below,
and used to check that
our result has the correct infrared pole structure.}

We can now present our result for the one-loop
MHV amplitude \eqref{loopint}\footnote{We thank Lance Dixon for 
pointing out some typos, both in these equations and elsewhere, in an 
earlier version of the paper.}:
\beqa
\nonumber
\cA_{\rm scalar} & = &
\cA_{\rm tree}\bigg\{
\sum_{m_1=j+1}^{i-1}\sum_{m_2=i+1}^{j-1}
{1\over 2} \big[ b_{m_1 m_2}^{ij} \big]^2
\,
B( q_{m_1, m_2 -1}^2 , q_{m_1+1,  m_2}^2,
q_{m_1+1,  m_2 -1}^2, q_{m_2+1, m_1 -1}^2)
\\ [6pt]\nonumber
& -  & \bigg(
 \,{8 \over 3}
\sum_{m_1=j+1}^{i-1}\sum_{m_2=i}^{j-1}
\Big[  \cA_{m_1 m_2}^{ij}\, T^{(3)} (m_1 , P , Q)
\, - \, (i \cdot j)  \tilde{\cA}^{ij}_{m_1 m_2}\,
T^{(2)} (m_1 , P , Q) \Big]
\\ [4pt]\nonumber
&+& 2
\sum_{m_1=j+1}^{i-1}\sum_{m_2=i}^{j-1}
\Big[  \cS^{ij}_{m_1 m_2}\, T^{(2)} (m_1 , P , Q)
\, + \,  \cI^{ij}_{m_1 m_2}\, T (m_1 , P , Q)
\Big]
\\[4pt]
 &&\qquad + \ ( i \longleftrightarrow j)\ \ \biggr) \
 \biggr\}
\ ,
\label{final}
\eeqa
where on the right hand side of \eqref{final}
a factor of $-4 \pi \l $ is understood,
where $\l$ is defined in \eqref{ubi}.
We can also introduce the coefficient
\beq
\label{art}
c_{m_1 m_2}^{i j} \ := \
{1\over 2}  \bigg[
{( i\, j \, m_2+1  \, m_1) \over \big( (m_2+1) \cdot m_1 \big)}
 \, - \,
{( i \, j \, m_2  \, m_1) \over (m_2 \cdot m_1 )}
\bigg]
\,
{ (i\, j\, m_1 \, Q) \, - \, ( i\, j\, Q\, m_1)
\over [(i \, + \, j)^2]^2}
\ ,
\eeq
which already appears
as the coefficient multiplying the triangle function $T$
in the $\cN \! = \! 1$ amplitude,
(see e.g.~Eq.~(2.19) of \cite{bbst}),
and rewrite \eqref{final} as
\beqa
\nonumber
\cA_{\rm scalar} & = &
\cA_{\rm tree}\bigg\{
\sum_{m_1=j+1}^{i-1}\sum_{m_2=i+1}^{j-1}
{1\over 2}
\big[ b_{m_1 m_2}^{ij} \big]^2
\,
B( q_{m_1, m_2 -1}^2 , q_{m_1+1 , m_2}^2,
q_{m_1+1 , m_2 -1}^2, q_{m_2+1, m_1 -1}^2)
\\ [6pt]\nonumber
& - &
\bigg(
\,
 {1\over 2}
\sum_{m_1=j+1}^{i-1}\sum_{m_2=i}^{j-1}
 {1\over 3} \, c^{ij}_{m_1 m_2}\,
\Big[ {
(i\, j\, m_1\,  Q) \,
(i\, j\,  Q \, m_1) \over 2 (i \cdot j )^2 }
\,  T^{(3)} (m_1 , P , Q)
\, +  \,\,
T  (m_1 , P , Q) \Big]
\\ [6pt]\nonumber
&+& 2
\sum_{m_1=j+1}^{i-1}\sum_{m_2=i}^{j-1}
\Big[  \cS^{ij}_{m_1 m_2}\, T^{(2)} (m_1 , P , Q)
\, + \,  \cI^{ij}_{m_1 m_2}\, T (m_1 , P , Q)
\Big]
\\[6pt]
&& \qquad +\ ( i \longleftrightarrow j) \ \ \biggr)\
 \biggr\}
\ .
\label{final2}
\eeqa
%
\begin{figure} [ht]
\label{scalartri1}
\vspace{.2in}
\centerline {
\includegraphics[width=4in]{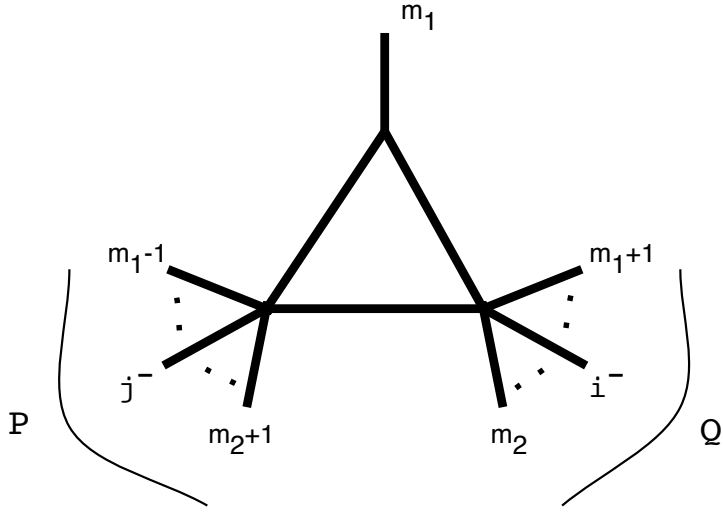}
}
\vspace{.2in}
\caption{\it One type of triangle function contributing to the amplitude
in the general case, where $i \in Q$, and $j \in P$.}
\end{figure}
\begin{figure} [ht]
\label{scalartri2}
\vspace{.2in}
\centerline {
\includegraphics[width=4in]{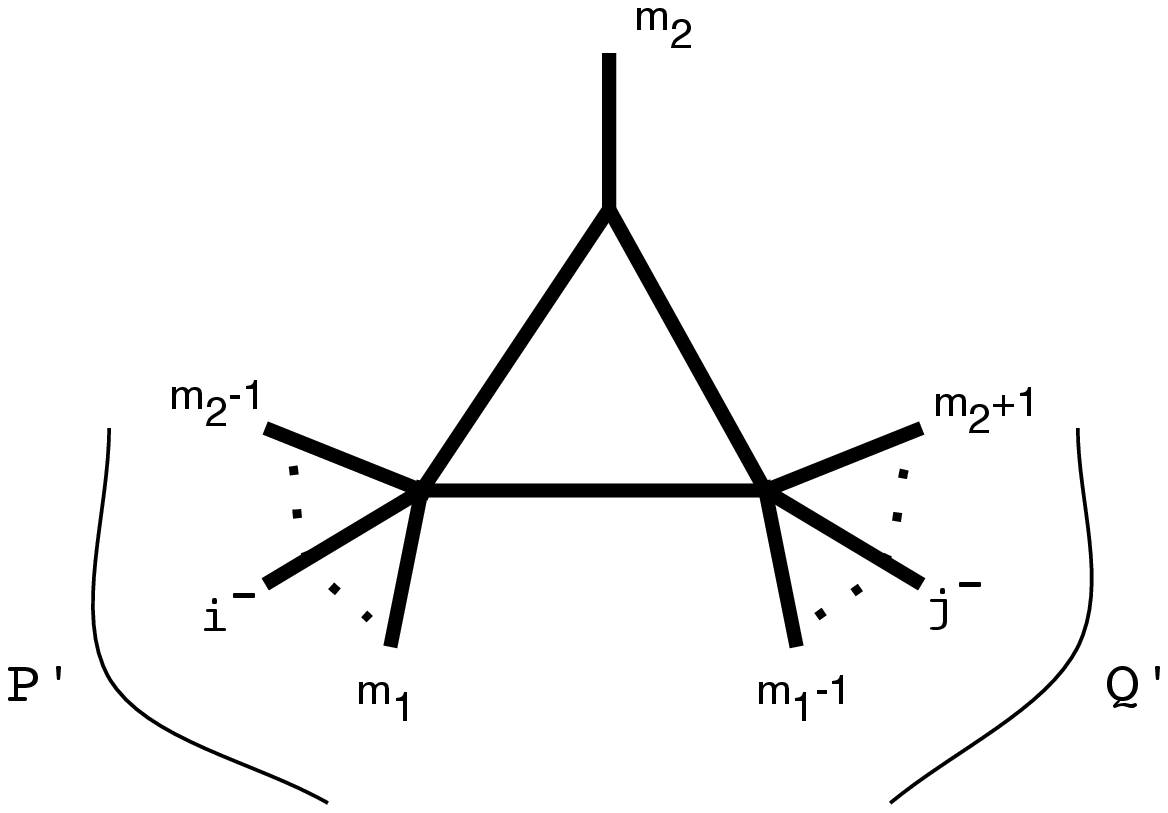}
}
\vspace{.2in}
\caption{\it Another type of triangle function contributing to the amplitude
in the general case.
By first shifting $m_1-1 \to m_1$,
and then swapping $m_1 \lrar m_2$,
we convert this into
a triangle function as in Figure 5 -- but with
$i$ and $j$ swapped. These are the triangle functions
responsible for the $i\lrar j$ swapped terms in
\eqref{final} -- or \eqref{final2}.
}
\end{figure}

Several remarks are in order.
\begin{itemize}
\item[{\bf 1.}]
As usual, the variables $q_{m_1, m_2 -1}^2$,
$q_{m_1+1 , m_2}^2$ correspond to the $s$- and $t$-channel
of the finite part of the ``easy two-mass'' box function with
massless legs $m_1$ and $m_2$, and massive legs
$q_{m_1+1 , m_2 -1}^2$, $q_{m_2+1, m_1 -1}^2$
(Figure 3).
\item[{\bf 2.}]
Compared to the range for $m_1$ and $m_2$
indicated in \eqref{range},
we have omitted $m_1 = i$ in the summation
of the triangles, as for this value the coefficients
$A$, $S$, $I$ defined in \eqref{alfredina}--\eqref{giuseppina}
vanish. Notice also that we have  $i \in Q$ and $j \in P$.
\item[{\bf 3.}]
In the case of adjacent negative-helicity gluons,
the only surviving terms are those  containing the
coefficient $c^{ij}_{m_1 m_2}$, on the
second line of \eqref{final} or \eqref{final2}.
We will return to this point in Section 5.
\item[{\bf 4.}]
We comment that, in contrast to the
adjacent case  (see \eqref{endresult}),
in the general case the $\cN=1$ chiral amplitude
does not separate out naturally in
the final result,  as one can see from
the coefficient of the box function
$B$ in \eqref{final}.
\end{itemize}

Next we wish to separate explicitly the infrared divergences
from \eqref{final}.
We can immediately anticipate that there will be four
infrared-divergent terms, corresponding to the four possible
degenerate triangles.
Two of these degenerate triangles occur when either
$P^2$ or $Q^2$ happen to vanish.
The other two
originate from the $i \lrar j$ swapped terms.

Let us consider first the terms arising from the summation with
 $i \lrar j$ unswapped.
When  $Q^2 =0$, it follows that $m_1 = i -1$ and $m_2 = i$
(see Figure 4).
When  $P^2 =0$, it follows that $m_1 = j+1 $ and $m_2 = j-1$
(see Figure 5).
Hence
\beqa
T^{(r)} (p, P, Q) & \to &
(-)^r \, {1\over \e} \, {(- t_{i-1} ^{[2]})^{-\e} \over
(t_{i-1}^{[2]})^r}
\ , \qquad Q^2 \to 0 \ ,
\\ \nonumber
T^{(r)} (p, P, Q) & \to &
- \, {1\over \e} \, {(- t_j^{[2]})^{-\e} \over (t_j^{[2]})^r}
\ , \qquad \quad \ \ P^2 \to 0 \ .
\eeqa
The infrared-divergent terms coming from $Q^2 =0$
are then easily extracted, and are
\beqa
\label{ir1}
- {1 \over 2 \e}  &\cdot & (- t_{i-1} ^{[2]})^{-\e} \,
4 (i \cdot j) \,
{(i\, j \, i-1 \, i+1) \over \big( (i+1) \cdot (i-1) \big)}
\\ \nonumber
&\cdot &
\bigg[
{8 \over 3} (i\cdot j)^2 \, - \,
2 \, {(i\, j \, i+1 \, i-1) \over \big( (i+1) \cdot (i-1) \big)}
(i \cdot j) \, + \,
 {(i\, j \, i+1 \, i-1) (i\, j \, i-1 \, i+1)
\over \big( (i+1) \cdot (i-1) \big)^2}
\bigg]
\ ,
\eeqa
and from $P^2 =0$
\beqa
\label{ir2}
- {1 \over 2 \e}  &\cdot & (- t_{j} ^{[2]})^{-\e} \,
4 (i \cdot j) \,
{(i\, j \, j-1 \, j+1) \over \big( (j+1) \cdot (j-1) \big)}
\\ \nonumber
&\cdot &
\bigg[
{8 \over 3} (i\cdot j)^2 \, - \,
2 \, {(i\, j \, j+1 \, j-1) \over \big( (j+1) \cdot (j-1) \big)}
(i \cdot j) \, + \,
 {(i\, j \, j+1 \, j-1) (i\, j \, j-1 \, j+1)
\over \big( (j+1) \cdot (j-1) \big)^2}
\bigg]
\ .
\eeqa
Likewise, from the ``swapped'' degenerate triangles we obtain
the following infrared-divergent terms:
\beqa
\label{ir3}
- {1 \over 2 \e}  &\cdot & (- t_{j-1} ^{[2]})^{-\e} \,
4 (i \cdot j) \,
{(i\, j \, j+1 \, j-1) \over \big( (j+1) \cdot (j-1) \big)}
\\ \nonumber
&\cdot &
\bigg[
{8 \over 3} (i\cdot j)^2 \, - \,
2 \, {(i\, j \, j-1 \, j+1) \over \big( (j+1) \cdot (j-1) \big)}
(i \cdot j) \, + \,
 {(i\, j \, j-1 \, j+1) (i\, j \, j+1 \, j-1)
\over \big( (j+1) \cdot (j-1) \big)^2}
\bigg]
\ ,
\eeqa
and
\beqa
\label{ir4}
- {1 \over 2 \e}  &\cdot & (- t_{i} ^{[2]})^{-\e} \,
4 (i \cdot j) \,
{(i\, j \, i+1 \, i-1) \over \big( (i+1) \cdot (i-1) \big)}
\\ \nonumber
&\cdot &
\bigg[
{8 \over 3} (i\cdot j)^2 \, - \,
2 \, {(i\, j \, i-1 \, i+ 1) \over \big( (i+1) \cdot (i- 1)
\big)}
(i \cdot j) \, + \,
 {(i\, j \, i-1 \, i+ 1) (i\, j \, i+1 \, i-1)
\over \big( (i+1) \cdot (i-1) \big)^2}
\bigg]
\ .
\eeqa


\subsection{Comments on twistor space interpretation}


We would like to make some brief comments on the interpretation in
twistor space of our result \eqref{final2}.
\begin{itemize}
\item[{\bf 1.}]
As noticed earlier, the coefficient $b^{ij}_{m_1 m_2}$ appears
already in the $\cN \! = \! 1$ chiral supermultiplet
contribution to a one-loop MHV amplitude, where it multiplies
the box function. It was noticed in
Section 4 of \cite{csw2} that $b^{ij}_{m_1 m_2}$
is a holomorphic function, hence it does not affect
the twistor space localisation of the finite box function.%
\footnote{We thank Dave Dunbar
for discussions on this point.}
\item[{\bf 2.}]
The coefficient $c^{ij}_{m_1 m_2}$ also
appears in the $\cN \! = \! 1$ amplitude, as the coefficient
of the triangles
(see e.g.~ Eq.~(2.19) of \cite{bbst}).
Its twistor space interpretation was considered in
Section 4 of \cite{csw2}, where it was found that
$c^{i j}_{m_1 m_2}$ has support on two lines in twistor space.
Furthermore, it was also found that the corresponding term
in the amplitude has a derivative of a delta function support
on coplanar configurations.
\item[{\bf 3.}]
The combination
$c^{ij}_{m_1 m_2}\,
(i\, j\, m_1\,  Q) \,
(i\, j\,  Q \, m_1) /  (i \cdot j )^2$
already appears in the case of adjacent negative-helicity gluons.
The localisation properties of the corresponding term in the
amplitude were considered in Section 5.3 of \cite{csw2},
and found to have, similarly to the previous case,
derivative of a delta function support
on coplanar configurations.
\item[{\bf 4.}]
On general grounds, we can argue that the remaining terms
in the amplitude have a twistor space interpretation
which is similar to that of the terms already considered.
The gluons whose momenta sum to  $P$
are contained on a line; likewise,   the gluons
whose momenta sum to $Q$ localise on another line.
\end{itemize}

We observe that the rational parts of the amplitude are not
generated from the MHV diagram construction
presented here.
Such rational terms were not present
for the $\cN=1$ and $\cN=4$ amplitudes
derived in \cite{bst,bbst,quig}.
However, for the amplitude studied here,
rational terms are required to ensure correct factorisation
properties \cite{Bern:1994cg}.


\section{Checks of the general result}

In this Section we present three consistency checks that we have
performed for the result \eqref{final}
(or \eqref{final2}) for the one-loop scalar contribution
to the MHV scattering amplitude. These checks are:
\begin{itemize}
\item[{ \bf 1.}]
For adjacent negative-helicity gluons,
the general expression \eqref{final} should reproduce the
previously calculated form \eqref{endresult}.
\item[{ \bf 2.}]
In the case of five gluons in the configuration $(1^-2^+3^-4^+5^+)$,
the  result \eqref{final}
should reproduce the known amplitude given in \cite{bdk9302280}.
\item[{\bf 3.}]
The result \eqref{final} should have
the correct infrared-pole structure.
\end{itemize}
We next discuss these requirements in turn.

\subsection{Adjacent case}
The amplitude where the
two negative-helicity external gluons are adjacent
is given in Section 7 of \cite{Bern:1994cg}
and was explicitly rederived in Section 3 of this paper
by combining MHV vertices, see
Eq.~\eqref{endresult}. It is easy to show
that our general result \eqref{final2} reproduce
correctly \eqref{endresult} as a special case.

To start with, recall that our result \eqref{final2}
is expressed in
terms of box-functions and triangle functions, see
Figure 3  and Figures 4, 5 respectively.
In the adjacent case,
the box functions are not present.
Indeed, in the sum \eqref{finbox} the negative-helicity gluons
can never be in adjacent positions (see Figure 3).

Next, we focus on the triangles of Figure 4.
In terms of these triangles, requiring
$i$ and $j$ to be adjacent eliminates the sum over $m_2$,
as we must have $m_2=i$ and $m_2 + 1 = j$.
Moreover, in this case $Q= q_{m_1 + 1, i}$, $P = q_{j, m_1 -1}$
and one has
\beqa
A_{ij}^{m_1 m_2} & = & -4 \, (i \cdot j)\ ,
\nonumber \\ \cr
S_{ij}^{m_1 m_2} & = & 0 \ , \qquad
I_{ij}^{m_1 m_2} \ = \ 0 \ ,
\eeqa
for $m_2=i$, and $m_2 + 1 = j$.
Similar simplifications occur for the swapped triangle.
Hence the only surviving terms
are those in the second line of \eqref{final}
(or \eqref{final2}), and it is then easy to see that
they generate the same amplitude
\eqref{atlast2} already calculated in
Section 3.

\subsection{Five-gluon amplitude}
The other special case is the non-adjacent
five-gluon amplitude $(1^-2^+3^-4^+5^+)$, given in
Eq.~(9) of \cite{bdk9302280}.
This amplitude may be written as
$c_{\Gamma} \cA_{\rm tree}$ times%
\footnote{The derivation in \cite{bdk9302280} used string-based
methods, which affects the coefficient of the pole term.
In \eqref{BDK}
we have written the pole coefficient
which matches the adjacent case.}
\begin{eqnarray}
 \label{BDK}\nonumber \\  \nonumber
 &&   \frac{1}{6\epsilon} - \frac{1}{6} \log(-s_{34})
   \ + \ \frac{\tofi^2\ \fito^2}{2^7(2\cdot 5)^4(1\cdot 3)^4}\
    B(s_{51}, s_{12}, 0, s_{34})
      \\ [8pt] \nonumber \cr
&& -\ \frac{1}{3}\ \frac{\tofi\ \fito}{2^4(2\cdot 5)(1\cdot 3)^4}
\ \left[
  \ \fito^2 \ \frac{\log(s_{12}/s_{34})}{(s_{12}-s_{34})^3}
    +\tofi^2\ \frac{\log(s_{34}/s_{51})}{(s_{34}-s_{51})^3}\ \right]
     \\ [8pt] \nonumber \cr
&&    +\ \frac{1}{3}\ \frac{1}{2^3(1\cdot 3)^3} \left[\ \foto\tofi^2
   \ \frac{\log(s_{34}/s_{51})}{(s_{34}-s_{51})^3}  \  \right]
    \\ [8pt] \nonumber \cr
&& - \ \frac{\tofi^2\ \fito^2}{2^6(2\cdot 5)^2(1\cdot 3)^4} \left[\
\  \frac{\log(s_{12}/s_{34})}{(s_{12}-s_{34})^2}
    - \frac{\log(s_{34}/s_{51})}{(s_{34}-s_{51})^2}\ \right]
\\ [8pt]\nonumber \cr
&&  + \ \frac{\tofi^2\ \fito^2}{2^6(2\cdot 5)^3(1\cdot 3)^4} \left[\
   \frac{\log(s_{12}/s_{34})}{(s_{12}-s_{34})}
    + \frac{\log(s_{34}/s_{51})}{(s_{34}-s_{51})}\ \right]
 \\ [8pt]\nonumber \cr
&&     -\ \frac{1}{3} \ \frac{1}{2^2(1\cdot 3)} \left[\ \tofi \
    \frac{\log(s_{34}/s_{51})}{(s_{34}-s_{51})}  \  \right]
  \\ [8pt] \cr
&&    \quad + \quad (1,4) \leftrightarrow (3,5)
\ \ ,
\end{eqnarray}
where the interchange on the last line applies
to all terms above it in this equation,
including the first two terms,
and the box function $B$ is defined in \eqref{boxf}.
In deriving this from  \cite{bdk9302280},
 we have used the dilogarithm identity
\begin{eqnarray}
 \label{spence} \nonumber
&& \!\!\! \!\!\! \!\!\! \!\!\!
  \Li(1-r) + \Li(1-s) + \log(r)\log(s)\  =
\Li\bigg(\frac{1-r}{s}\bigg) +
\Li\bigg(\frac{1-s}{r}\bigg)
-  \Li\bigg(\frac{1-s}{r}\frac{1-r}{s}\bigg).
\\ \cr  &&
\end{eqnarray}
We have checked explicitly that our expression for the
$n$-gluon non-adjacent amplitude \eqref{final},
when specialised to the case with five gluons
in the configuration $(1^-2^+3^-4^+5^+)$,
yields precisely the result
\eqref{BDK} above.
For the terms involving dilogarithms, this is
easily done. For the remaining terms,
which contain logarithms, a more involved calculation
is necessary using various spinor identities
from Appendix D.  A straightforward method of
doing this calculation begins
with the explicit sum over MHV diagrams in this case,
isolating the coefficients of each logarithmic
function such as e.g.~$\log(s_{12})$,
and then checking that these coefficients
match those in \eqref{BDK}.
The remaining $1/\epsilon$ term arises
from the following discussion.


\subsection{Infrared-pole structure}

The infrared-divergent terms (poles in $ 1 / \e$)
can easily be extracted  from \eqref{ir1}--\eqref{ir4}
by simply replacing $ ( - t_r^{[2]})^{- \e} \to 1$
($r = i-1,  i,  j-1,  j$).
Consider first the terms in \eqref{ir2} and \eqref{ir3}.
After a little algebra, and using
\beq
(i \, j \, j+1 \, j-1) \, + \, (i \, j \, j-1 \, i-1)
\ = \
4\,  (i \cdot j) \, \big( (j-1) \cdot (j +1) \big)
\ ,
\eeq
one finds that these two contributions
add up to
\beq
- {64 \over 3\, \e} \ (i \cdot j)^4
\ .
\eeq
Similarly, the pole contribution arising from
\eqref{ir1} and \eqref{ir4} gives an additional contribution
of  $ - ({64 /  3\, \e})\  (i \cdot j)^4$.
Reinstating a factor  of
$-2 \cdot 2^{-8} (i\cdot j)^{-4} \cdot  \cA_{\rm tree}$, we see that
the pole part of \eqref{final} is simply given by
\beq
\left. \cA_{\rm scalar}\right|_{\rm \e-pole}
 \ = \ {\cA_{\rm tree} \over 3}
\ .
\eeq
Hence our result \eqref{final} has the expected
infrared-singular behaviour.


\section*{Acknowledgements}

It is a pleasure to thank Lance Dixon, Dave Dunbar,
Michael Green, Marek Karliner, Valya Khoze,
David Kosower, Marco Matone and Sanjaye Ramgoolam
for discussions.
GT acknowledges the support of PPARC.

\newpage


\startappendix


\Appendix{Passarino-Veltman reduction}

%
In Section 2  we saw that a typical term in the
cut-constructible part of the Yang-Mills
amplitude is the dispersion integral
of the following phase space integral:
\begin{eqnarray}
\label{partintegral}
\cC (m)  \, := \,
\int\! d{\rm LIPS}(l_2, -l_1; P_{L;z}) \,
\frac{
 {\tr}_{+}(\kslash_{1}\, \kslash_{2}\,
\Pslash_{L;z} \, \lslash_{2})
\, {\tr}_{+}(\kslash_{1}\, \kslash_{2}
\, \lslash_{2} \, \Pslash_{L;z})
\, {\tr}_{+}(\kslash_1 \, \kslash_2\,
\kslash_{m} \, \lslash_2)
}
{(l_2\cdot m) \, (k_1\cdot k_2)^3 \, (P_{L;z}^2)^2
}
\, .
\end{eqnarray}
The goal of this Appendix is to perform the
Passarino-Veltman reduction \cite{pv} of
\eqref{partintegral}.
To this end, we rewrite $\cC (m ) $ as
\beq
\label{partintegralii}
\cC (m)  \ = \
{
{\tr}_{+}(\kslash_{1}\, \kslash_{2}\,
\Pslash_{L;z} \, \gamma_{\mu} )
\ {\tr}_{+}(\kslash_{1}\, \kslash_{2}
\, \gamma_\nu \, \Pslash_{L;z})
\
{\tr}_{+}(\kslash_1 \, \kslash_2\,
\kslash_{m} \, \gamma_{\rho})
\over
 (k_1\cdot k_2)^3 (P_{L;z}^2)^2 \,
}
\ \cI^{\m \n \r} (m , P_{L;z})
\ ,
\eeq
where%
\footnote{For the rest of this Appendix we drop the subscript
$z$ in $P_{L;z}$ for the sake of brevity.}
\beq
\label{imunu}
\cI^{\m \n \r} (m, P_{L})
\ = \
\int\! d{\rm LIPS}(l_2,-l_1;P_{L})
\,
\frac{l_2^{\mu}\, l_2^{\nu} \, l_2^{\r} }{
(l_2\cdot m)}
\ .
\eeq
On general grounds, $\cI^{\mu\nu\r} (m, P_{L})$
can be decomposed as
\begin{eqnarray}
\label{ansatz}
\cI^{\mu\nu\r} &=&
m^{\mu}m^{\nu}m^{\r} \, \cI_1 \ + \
(m^{\mu}m^{\nu}P_{L}^{\r}
\, + \,
m^{\mu}P_L^{\nu}m^{\r} \, + \,
P_L^{\mu}m^{\nu}m^{\r}
) \, \cI_2
\nonumber\\ \cr
&+ &
(m^{\mu}P_L^{\n}P_L^{\r}
\, + \,
P_L^{\mu}m^{\nu}P_L^{\r}
+
P_L^{\mu}P_L^{\nu}m^{\r} )
\, \cI_3
\ + \
P_L^{\mu}P_L^{\nu}P_L^{\r}\, \cI_4
\nonumber\\ \cr
&+ &
( \eta^{\m \n} m^{\r}
\, + \, \eta^{\m \r} m^{\n} \, + \, \eta^{\n \r} m^\m)\,
\cI_5
\ + \
(\eta^{\m \n} P_L^{\r}
\, + \, \eta^{\m \r} P_L^{\n} \, + \, \eta^{\n \r} P_L^\m)
\cI_6
\ ,
\end{eqnarray}
for some coefficients $\cI_i, i=0,...,6$. One can
then contract with different combinations of
the independent momenta in order to solve for the
$\cI_{i}$. Introducing the quantities
\beqa
\label{pvstuff}
A & := &  \ m_\m m_\n m_\r \, I^{\m \n \r}\ ,
\cr
B & := &  \ m_\m m_\n P_{L\r} \, I^{\m \n \r}\ ,
\cr
C & := &  \ m_\m P_{L\n} P_{L\r} \, I^{\m \n \r}\ ,
\cr
D & := &  \ P_{L\m} P_{L\n} P_{L\r} \, I^{\m \n \r}\ ,
\cr
E & := &  \ \eta_{\m \n} m_{\r} \, I^{\m \n \r}\ \ = \ 0
\ ,
\cr
F & := &  \ \eta_{\m \n} P_{L\r} \, I^{\m \n \r}\ = \ 0 \ ,
\eeqa
the result for the Passarino-Veltman reduction of
$\{\cI_{1},\ldots ,  \cI_{6}\}$
in the basis $\{A, \ldots , D \}$ is:
\beqa
\label{basis}
\cI_2 & = & \left( 5 (P_L^2)^2 /
\big( 2 (m \cdot P_L)^5\big)\, , \,
-6 P_L^2 / (m \cdot P_L)^4\,  , \,
3  / (m \cdot P_L)^3 \, ,\,  0\right)
\ ,
\cr
 \cI_3 & = & \left( -2 P_L^2 /  (m \cdot P_L)^4\, , \,
3/ (m \cdot P_L)^3 \, ,\,  0\, ,\,  0\right)
\ ,
\cr
\cI_4& = & \left( 1 / (m \cdot P_L)^3\, , \,
0 \, ,\,  0\, , \, 0\right)
\cr
\cI_5 & = & \left( - (P_L^2)^2 /  \big( 2 (m \cdot P_L)^4\big)
\, ,
3 P_L^2 / \big( 2(m \cdot P_L)^3\big) \, , \,
-1  / (m \cdot P_L)^2\, ,\,  0\right)
\ ,
\cr
\cI_6& = & \left( P_L^2 / \big(2 (m \cdot P_L)^3)\, ,
-1/(m \cdot P_L)^2 \, , 0\, , 0\right)
\ .
\eeqa
We omit the decomposition for $\cI_1$,
as the corresponding term in
\eqref{ansatz} drops out of all future expressions due to
$k_m^2 =0$.

Finally, using the methods of \cite{bbst} and the results of Appendix C,
the integrals in (\ref{pvstuff}) are found to be,
keeping only terms to  order $\mathcal{O}(\epsilon^0)$,
\beqa
\label{explicit}
A \ &=& \ (m\cdot P_L)^2\, {4 \over 3} \pi \lambda \ ,
\\ [2pt]
\cr
B \ &=& \ P_L^2(m\cdot P_L)
\,
\pi \lambda \, ,\\ [2pt]
\cr
C \ &=& \ (P_L^2)^2\,
\pi \lambda \, ,\\ [2pt]
\cr
D \ &=& \ -\frac{(P_L^2)^3}{8(m\cdot P_L)}\, \frac{4\pi}{\epsilon}\,
\lambda \, ,
\eeqa
where
\beq
\l \ :=  \frac { \pi^{ \frac{1}{2} - \epsilon }
 } {  4^{1-\e} \, \Gamma \big(\frac{1}{2} - \epsilon\big) }
\ .
\eeq
%


\Appendix{Evaluating the integral of ${\cal C}(a,b)$}


The basic expression which arises in
the MHV diagram construction in this paper is
\beqa
\label{basicintegrand}
{\cal C}(a,b) = \frac{\langle i \, l_1 \rangle \,
\langle j \, l_1 \rangle^2 \,
\langle i \, l_2 \rangle^2 \,
\langle j \, l_2 \rangle}{\langle i \, j \rangle^4 \,
\langle l_1 \, l_2 \rangle^2}
\frac{\langle i \, a \rangle \, \langle j \, b \rangle}
{\langle l_1 \, a \rangle \, \langle l_2 \, b \rangle}.
\eeqa
We wish to integrate this expression
over the Lorentz invariant phase space. We begin by simplifying
it, using multiple applications of the Schouten identity.
First note that using this identity twice, one deduces that
\begin{eqnarray}
\label{basicmanip} \frac{\lrangle{i}{l_2} \, \lrangle {j}{l_1} }
                     {\lrangle{l_1}{a}\,\lrangle{l_2}{b} }
\lrangle{a}{b}^2
     &=& \lrangle{i}{a}\lrangle{b}{j} +\lrangle{i}{a}\lrangle{a}{j}
     \frac{\lrangle{l_1}{b}}{\lrangle{a}{l_1}} +\lrangle{b}{j}
\lrangle{i}{b}
     \frac{\lrangle{l_2}{a}}{\lrangle{b}{l_2}}
\\ \nonumber \cr
     &+&  \lrangle{a}{j}\lrangle{i}{b}
     -\lrangle{a}{j}\lrangle{i}{b}
\frac{\lrangle{l_1}{l_2}\lrangle{a}{b}}{\lrangle{a}{l_1}
     \lrangle{b}{l_2}}
\ .
\end{eqnarray}
Now use this identity in ${\cal C}(a,b)$.
This generates five terms, which we will label
(in correspondence with the ordering arising from the order of
terms in \eqref{basicmanip} above)
as $T_i, i=1,\dots,4$, and $U$.
The $T_i$ have dependence on the loop momenta such that
we may use the phase space integrals
of Appendix C to calculate them. The term $U$
is more complicated; however,
one may again use the identity \eqref{basicmanip},
generating another five terms,
which we will label $T_5,\dots,T_8$, and $V$.
Again, the expressions in $T_i, i=4,...,8$ may be calculated using
the integrals of Appendix C.
Finally, the term $V$ may be simplified,
here using the identity \eqref{basicmanip}
with $i$ and $j$ interchanged.
This generates a further five terms, which we label
$T_9,\dots,T_{13}$.
The explicit forms of these terms follow:
\begin{eqnarray}
\label{allterms1}
 T_1& =&
\frac{\ijba^2\ijlolt\ijltlo}{2^{8}(i\cdot j)^4
(a\cdot b)^2(l_1\cdot l_2)^2} \ ,
\\  [2pt]\cr
\label{T1}
 T_2& =&
\frac{\ijab\ijba\ijlolt\ijltlo\ibloa}{2^{10}(i\cdot j)^4
(a\cdot b)^2(l_1\cdot l_2)^2(i\cdot b)(a\cdot l_1)} \ ,
\\ [2pt]\cr
\label{T2}
 T_3& =&
\frac{\ijab\ijba\ijlolt\ijltlo\jaltb}{2^{10}(i\cdot j)^4
(a\cdot b)^2(l_1\cdot l_2)^2(j\cdot a)(b\cdot l_2)} \ ,
\\ [2pt]\cr
\label{T3}
T_4& =& -\frac{\ijab\ijba\ijlolt\ijltlo}{2^{8}(i\cdot j)^4
(a\cdot b)^2(l_1\cdot l_2)^2} \ ,
\end{eqnarray}
and
\begin{eqnarray}
\label{allterms2}
T_5& =& \frac{\ijba^2\ijab\ijltlo}{2^{8}(i\cdot j)^4(a\cdot b)^3
(l_1\cdot l_2)}\ ,
\\ [2pt]\cr
\label{T5}
T_6& =& \frac{\ijab^2\ijba\ijltlo\ibloa}
{2^{10}(i\cdot j)^4(a\cdot b)^3(l_1\cdot l_2)(i\cdot b)
(a\cdot l_1)}
\ ,
\\ [2pt]\cr
\label{T6}
T_7& =& - \frac{\ijab\ijba^2\ijltlo\ialtb}{2^{10}(i\cdot j)^4
(a\cdot b)^3(l_1\cdot l_2)(i\cdot a)(b\cdot l_2)}
\ ,
\\ [2pt]\cr
\label{T7}
T_8& =& -\frac{\ijab^2\ijba\ijltlo}{2^{8}(i\cdot j)^4
(a\cdot b)^3(l_1\cdot l_2)}
\ ,
\end{eqnarray}
and
\begin{eqnarray}
\label{allterms3}
T_{9}& =& \frac{\ijab^3\ijba}{2^{8}(i\cdot j)^4(a\cdot b)^4}\ ,
\\ [2pt]\cr
\label{T9}
T_{10}& =& \frac{\ijab^2\ijba^2
\jbloa}{2^{10}(i\cdot j)^4
(a\cdot b)^4(j\cdot b)(a\cdot l_1)}
\ ,
\\ [2pt]\cr
\label{T10}
T_{11}& =&  \frac{\ijab^2\ijba^2\ialtb}{2^{10}(i\cdot j)^4
(a\cdot b)^4(i\cdot a)(b\cdot l_2)}
\ ,
\\ [2pt]\cr
\label{T11}
T_{12}& =& -\frac{\ijab^2\ijba^2}{2^{8}(i\cdot j)^4(a\cdot b)^4}
\ ,
\\ [2pt]\cr
\label{T12}
T_{13}& =& \frac{\ijba^2\ijab^2\bltloa}{2^{10}(i\cdot j)^4
(a\cdot b)^4(a\cdot l_1)(b\cdot l_2)}
\ ,
\end{eqnarray}
The expression ${\cal C}(a,b)$ is
then the sum of the terms $T_i, i=1,\dots,13$.

Before performing the phase space integrals,
it proves convenient to
%
%
collect the resulting expressions in pairs as $T_1+T_2$,
$T_3+T_4$, $T_5+T_6$,
$T_7+T_8$,  $T_9+T_{11}$ and
$T_{10}+T_{12}$,
we are led to the following decomposition:
\beqa
-\cC (a, b) & = &
{ \ijlolt \ijltlo\ijloa\ijblt \over
2^8 (i \cdot j)^4 (l_1 \cdot l_2 )^2 (l_1 \cdot a)
(l_2 \cdot b) }
\nonumber \\ \cr
&=&
{1 \over 2^8 (i \cdot j)^4} \,
(\cH_1 \, + \, \cdots \, + \cH_4 )
\ ,
\eeqa
where
 \beqa \nonumber
\!\!\!\!\!\!\!\!\cH_1 &:=&
{ \ijba  \ijlolt \ijltlo \over (l_1 \cdot l_2)^2 \, (a \cdot b)}
\bigg[
{\ijloa \over (l_1 \cdot a)} -
{\ijltb \over (l_2 \cdot b)}
\bigg]
\ ,
\\ [5pt]\nonumber \cr
\!\!\!\!\!\!\!\! \cH_2 &:=&
{ \ijab \ijba  \ijltlo \over (l_1 \cdot l_2) \, (a \cdot b)^2}
\bigg[
{\ijloa \over (l_1 \cdot a)} -
{\ijltb \over (l_2 \cdot b)}
\bigg]
\ ,
\\ [5pt]\nonumber \cr
\!\!\!\!\!\!\!\! \cH_3 &:=&
- { ( \ijab )^2 \ijba  \over  (a \cdot b)^3 }
\bigg[
{\ijloa \over (l_1 \cdot a)} -
{\ijltb \over (l_2 \cdot b)}
\bigg]
\ ,
\\ [5pt] \cr
\!\!\!\!\!\!\!\! \cH_4 &:=&
 { ( \ijab )^2 (\ijba )^2 \loablt  \over  4 (a \cdot b)^4
\, (l_1 \cdot a) \, (l_2 \cdot b) }
\ .
\eeqa

Finally, we perform the phase space integrals of
the above expressions, using the formulae  in
Appendix C below. One finds quickly
that the divergent (as $\epsilon\rightarrow 0$) part
of the total expression is zero.
The finite part, after further spinor
manipulations, becomes the expression
we have given in \eqref{pspaceresult}.


\Appendix{Phase space integrals}


The basic method which we use for evaluating
Lorentz-invariant phase space integrals has been
outlined in our earlier papers \cite{bst, bbst}.
Here we will just quote the results which we need.
In the following we will use a shorthand notation where
$\int \equiv
\int \!d^{4-2\epsilon}{\rm LIPS} (l_2 , -l_1 ; P_{L})$,
and a common factor of $ 4 \pi \l ( -P_{L;z})^2 $
is understood to multiply all expressions, where
$\l$ is the ubiquitous factor
\beq
\label{ubi}
\l \ :=  \frac { \pi^{ \frac{1}{2} - \epsilon }
 } {  4^{1-\e} \, \Gamma \big(\frac{1}{2} - \epsilon\big) }
\ .
\eeq
In the following we define
$\alpha = (a\cdot P)$, $\beta=(b\cdot P)$,
$N(P) = (a\cdot b)P^2 -2(a\cdot P)(b\cdot P)$
and drop the $L;z$ subscripts
on $P_{L;z}$ for clarity.

Firstly we quote the results from Appendix B of
\cite{bbst} up terms of order
$\mathcal{O}(\epsilon^0)$:
\begin{eqnarray}
\label{beebop1}
\int 1 &=& 1
\ , \qquad
\int \frac{1}{(a\cdot l_1)}= - \frac{1}{\epsilon\alpha}\ ,  \qquad
\int \frac{1}{(b\cdot l_2)} = \frac{1}{\epsilon\beta}
\ ,
\\ [6pt]\nonumber
\label{beebop4}
&&\int \frac{1}{(a\cdot l_1)(b\cdot l_2)}
=
-\frac{4}{N(P)}\left(\frac{1}{\epsilon}+ L\right)\ ,
\nonumber
\end{eqnarray}
where
$$
L = \log\left( 1 - \frac{(a\cdot b)}{N}P^2 \right).
$$
%
From this, we can derive recursively the following
integrals (up to $\cO (\e^0)$):
\begin{eqnarray}
\label{beebop5}
\int l_1^\mu &=&  \frac{1}{2}P^\mu, \qquad
\int l_2^\mu \ = \
- \,  \frac{1}{2}P^\mu
\ ,
\\ [4pt]\nonumber
\label{beebop6}
\int l_1^\mu l_1^\nu &=&
\int l_2^\mu l_2^\nu \ =\   \frac{1}{3}
\left( P^\mu P^\nu \, - \,
 \frac{1}{4}\eta^{\mu\nu}P^2\right)
\ ,
\\ [4pt]\nonumber
\label{beebop8}
\int \frac{l_1^\mu}{(a\cdot l_1)}
&=&
-\, \frac{P^2}{2\epsilon\alpha^2}a^\mu \, + \,
\frac{1}{\alpha}P^\mu \, - \, \frac{P^2}{\alpha^2}a^\mu
\ ,
\\ [4pt]\nonumber
\label{beebop9}
\int \frac{l_2^\mu}{(b\cdot l_2)}
&=&
-\, \frac{P^2}{2\epsilon\beta^2}b^\mu \, +
\, \frac{1}{\beta}P^\mu \, - \,
\frac{P^2}{\beta^2}b^\mu
\ ,
\nonumber
\end{eqnarray}
and
\begin{eqnarray}
\label{beebop10}
\int \frac{l_1^\mu l_1^\nu}{(a\cdot l_1)} &=&
-\frac{P^4}{4\epsilon\alpha^3}a^\mu a^\nu
+\frac{1}{2\alpha}P^\mu P^\nu +
\frac{P^2}{2\alpha^2} P^{(\mu}a^{\nu)}
-\frac{3P^4}{4\alpha^3}a^\mu a^\nu -
\frac{P^2}{4\alpha}\eta^{\mu\nu}
\ ,
\\ [4pt]\nonumber
\label{beebop11}
\int \frac{l_2^\mu l_2^\nu}{(b\cdot l_2)} &=&
\frac{P^4}{4\epsilon\beta^3}b^\mu b^\nu
-\frac{1}{2\beta}P^\mu P^\nu -
\frac{P^2}{2\beta^2} P^{(\mu}b^{\nu)}
+\frac{3P^4}{4\beta^3}b^\mu b^\nu +
\frac{P^2}{4\beta}\eta^{\mu\nu}
\ ,
\\ [4pt]\nonumber
\label{beebop12}
\int \frac{l_2^\mu }{(a\cdot l_1)(b\cdot l_2)} &=&
\frac{1}{\epsilon N}\left(
2P^\mu -\frac{P^2}{\alpha}a^\mu +
\frac{P^2}{\beta}b^\mu \right)
+ \frac{2L}{N} \left( P^\mu - \frac{\beta}{(a\cdot b)}a^\mu +
\frac{\alpha}{(a\cdot b)}b^\mu \right)
\, .
\nonumber
\end{eqnarray}
Finally, there are integrals involving
cubic powers of loop momenta in the numerator.
The first is
\begin{eqnarray}
\label{cubic2}
\!\!\!\!\!\!\!\!\!\!\!\!\!\!\int
\frac{l_1^\mu l_1^\nu l_1^\rho}{(a\cdot l_1)}\!\!& =&\!\!
 \frac{P^4}{4\alpha^3} P^{(\mu}a^\nu a^{\rho)}  +
\frac{P^2}{4\alpha^2} P^{(\mu}P^\nu a^{\rho)}
 + \frac{1}{3\alpha} P^\mu P^\nu P^{\rho} -
\frac{P^4}{8\alpha^2} \eta^{(\mu\nu} a^{\rho)}
 - \frac{P^2}{4\alpha} \eta^{(\mu\nu} P^{\rho)}
\, ,
\end{eqnarray}
where we have suppressed terms cubic in $a$
as they prove not to contribute
when this integral
is contracted into the products of
Dirac traces which appear in the expressions in
Appendix B.
The second cubic integral required is
\begin{eqnarray}
\label{cubic1}
\!\!\!\!\!\!\!\!\!\!\!\!\!\!\int
\frac{l_2^\mu l_2^\nu l_2^\rho}{(b\cdot l_2)}\!\!& =&\!\!
 \frac{P^4}{4\b^3} P^{(\mu}b^\nu b^{\rho)}  +
\frac{P^2}{4\b^2} P^{(\mu}P^\nu b^{\rho)}
 + \frac{1}{3\b} P^\mu P^\nu P^{\rho} -
\frac{P^4}{8\b^2} \eta^{(\mu\nu} b^{\rho)}
 - \frac{P^2}{4\b} \eta^{(\mu\nu} P^{\rho)}
\, ,
\end{eqnarray}
again suppressing  terms cubic in $b$ which will not contribute.

%
%


\Appendix{Spinor identities}


We collect here some formulae useful for the calculations
presented in this paper. The Schouten identity is
\beq
\label{schouten}
\langle i\, j\rangle\langle k\, l\rangle =
\langle i\, k\rangle\langle j\, l\rangle + \langle i\,
l\rangle\langle k\, j\rangle \ .
\eeq
Other identities are:
\begin{eqnarray}
\label{spinor1}
\left[ i\, j\right] \lan j\,
i\ran
&=&
{\tr}_{+}(\kslash_i \kslash_j)=2(k_i\cdot k_j)
\ ,
\\ \cr
\label{spinor2}
\left[  i\, j\right] \lan j\, l\ran
\left[  l\,
m\right] \lan m\, i\ran
&=&
{\tr}_{+}(\kslash_i \kslash_j
\kslash_{l} \kslash_m)
\ ,
\\ \cr
\label{spinor3}
\left[ i\, j\right]
\lan j\, l\ran \left[  l\,
m\right] \lan m\, n\ran \left[  n\, p\right]
\lan p\, i\ran
&=&
{\tr}_{+}(\kslash_i \kslash_j \kslash_{l} \kslash_m \kslash_n
\kslash_p)\ ,
\end{eqnarray}
for momenta
$k_i,k_j,k_l,k_m,k_n,k_p$.
We also have, for null momenta $i,j,k,a,b,$
\beq
\label{simplifier}
\frac{\ijab\jakb}{(j\cdot a)} = - \frac{\ijba\iakb}{(i\cdot a)}
\ .
\eeq
For dealing with Dirac traces,
we have the following identities%
\footnote{The appearance of a Greek letter such as $\mu$ inside a
trace indicates that the relevant
gamma matrix is to be inserted at that point.}
\begin{eqnarray}
\label{trace1}
{\tr}_{+}(\kslash_i \kslash_j \kslash_{l}
\kslash_m)
&=&
{\tr}_{+}(\kslash_m \kslash_l \kslash_{j}
\kslash_i) \ = \
{\tr}_{+}(\kslash_l \kslash_m \kslash_{i}\kslash_{j})
\ ,
\\ \cr
\label{trace2}
{\tr}_{+}(\kslash_i \kslash_j \kslash_{l}
\kslash_m)
&=&
4(k_i\cdot k_j)(k_l\cdot k_m)\, -\, {\tr}_{+}(\kslash_j
\kslash_i \kslash_{l} \kslash_m)
\ ,
\\ \cr
\label{trace3}
\ijmuP \, \ijmum &=& 0
\ ,
\\ \cr
\label{trace4}
\ijmuP \, \ijmmu &=& 4(i\cdot j)\, \ijmP
\ .
\end{eqnarray}
%



\newpage
\begin{thebibliography}{99}

\bibitem{witten} E.~Witten,
{\it Perturbative gauge theory as a string theory in twistor
space}, {\tt hep-th/0312171}.

\bibitem{csw} F.~Cachazo, P.~Svrcek and E.~Witten,
{\it MHV vertices and tree amplitudes in gauge theory}, JHEP {\bf 0409} (2004) 006,
{\tt hep-th/0403047}.

\bibitem{vvkreview}
V.V. Khoze, {\it Gauge Theory Amplitudes, Scalar Graphs and Twistor Space},
To appear in {\it From Fields to Strings: Circumnavigating Theoretical Physics},
in memory of Ian Kogan,
{\tt hep-th/0408233}.

\bibitem{BerkWitt2} N. Berkovits and E. Witten, {\it Conformal Supergravity
in Twistor-String Theory}, JHEP {\bf 0408} (2004) 009,
{\tt hep-th/0406051}.

\bibitem{csw2}
F.~Cachazo, P.~Svrcek and E.~Witten,
{\it Twistor space structure of one loop amplitudes in gauge theory},
JHEP {\bf 0410} (2004) 074,
{\tt hep-th/0406177}.

\bibitem{bst}
A.~Brandhuber, B.~Spence and G.~Travaglini,
{\it One-Loop Gauge Theory
Amplitudes in N=4 super Yang-Mills from MHV Vertices},
Nucl.\ Phys.\ B {\bf 706}, 150 (2005),
{\tt  hep-th/0407214}.

\bibitem{csw3}
F. Cachazo, P. Svrcek and E. Witten,
{\it Gauge Theory Amplitudes In Twistor Space And Holomorphic
Anomaly}, JHEP {\bf 0410} (2004) 077,
{\tt hep-th/0409245}.

\bibitem{benabern}
I.~Bena, Z.~Bern, D.~A.~Kosower and R.~Roiban,
{\it Loops in Twistor Space},
{\tt hep-th/0410054}

\bibitem{freddy3}
R.~Britto, F.~Cachazo, B.~Feng,
{\it Coplanarity in Twistor Space of $N=4$ Next-To-MHV
One-Loop Amplitude Coefficients},
{\tt hep-th/0411107}.

\bibitem{freddy}
F.~Cachazo, {\it
Holomorphic Anomaly Of Unitarity Cuts And One-Loop Gauge Theory Amplitudes},
{\tt hep-th/0410077}.

\bibitem{freddy2}
R.~Britto, F.~Cachazo, B.~Feng,
{\it Computing One-Loop Amplitudes from the
Holomorphic Anomaly of Unitary Cuts},
{\tt hep-th/0410179}.

\bibitem{new}
Z.~Bern, V.~Del Duca, L.~J.~Dixon, and D.~A.~Kosower,
{\it
All Non-Maximally-Helicity-Violating One-Loop
Seven-Gluon Amplitudes in N=4 Super Yang-Mills Theory},
{\tt hep-th/0410224}.

\bibitem{BBDD}
S.~J.~Bidder, N.~E.~J.~Bjerrum-Bohr, L.~J.~Dixon and
D.~C.~Dunbar,
{\it $N=1$ Supersymmetric One-loop Amplitudes and the Holomorphic Anomaly of
Unitarity Cuts}, {\tt hep-th/0410296}.

\bibitem{BBDP}
S.~J.~Bidder, N.~E.~J.~Bjerrum-Bohr, D.~C.~Dunbar and
W.~B.~Perkins,
{\it Twistor Space Structure of the Box Coefficients of $N=1$ One-loop
Amplitudes}, {\tt hep-th/0412023}.

\bibitem{bbst}
J.~Bedford, A.~Brandhuber, B.~Spence and G.~Travaglini, {\it A Twistor
Approach to One-Loop Amplitudes in ${\cal N} \! = \! 1$ Supersymmetric
Yang-Mills Theory},
Nucl.\ Phys.\ B {\bf 706}, 100 (2005),
{\tt hep-th/0410280}.

\bibitem{quig}
C.~Quigley and M.~Rozali, {\it One-Loop MHV Amplitudes in Supersymmetric
Gauge Theories}, {\tt hep-th/0410278}.



\bibitem{Bern:zx}
Z.~Bern, L.~J.~Dixon, D.~C.~Dunbar and D.~A.~Kosower, {\it One
Loop N Point Gauge Theory Amplitudes, Unitarity And Collinear
Limits,} Nucl.\ Phys.\ B {\bf 425} (1994) 217, {\tt
hep-ph/9403226}.



\bibitem{higgs}
L.~J.~Dixon, E.~W.~N.~Glover and V.~V.~Khoze,
{\it MHV Rules for Higgs Plus Multi-Gluon Amplitudes}, JHEP {\bf 0412} (2004) 015,
{\tt hep-th/0411092}.

\bibitem{Bern:1994cg}
Z.~Bern, L.~J.~Dixon, D.~C.~Dunbar and D.~A.~Kosower, {\it Fusing
gauge theory tree amplitudes into loop amplitudes,} Nucl.\ Phys.\
B {\bf 435} (1995) 59, {\tt hep-ph/9409265}.

\bibitem{bdk9302280}
Z.~Bern, L.~J.~Dixon and D.~A.~Kosower, {\it One-Loop Corrections to Five-Gluon
amplitudes,} Phys. Rev. Lett. {\bf 70} (1993) 2677-2680,
{\tt hep-ph/9302280}.

\bibitem{dixon}
L.~J.~Dixon, {\it Calculating scattering amplitudes efficiently,}
TASI Lectures 1995, {\tt hep-ph/9601359}.


\bibitem{rsv}
R.~Roiban, M.~Spradlin and A.~Volovich, {\it A googly amplitude
from the B-model in twistor space}, JHEP {\bf 0404} (2004) 012,
{\tt hep-th/0402016.}


\bibitem{Berkovits} N. Berkovits, {\it
An Alternative String Theory in Twistor Space for $N=4$
Super-Yang-Mills}, {\tt hep-th/0402045}.

\bibitem{rv}
R.~Roiban and A.~Volovich, {\it All googly amplitudes from the
B-model in twistor space}, {\tt hep-th/0402121}.

\bibitem{rsv2}
R.~Roiban, M.~Spradlin and A.~Volovich, {\it On the tree-level
S-matrix of Yang-Mills theory},
Phys.\ Rev.\ D {\bf 70} (2004) 026009,
{\tt hep-th/0403190.}

\bibitem{BerkMotl} N. Berkovits and L. Motl, {\it Cubic Twistorial String Field
Theory}, J. High Energy Phys. {\bf 0404} (2004) 056, {\tt
hep-th/0403187}.

\bibitem{gmn} S. Gukov, L. Motl and  A. Neitzke, {\it Equivalence of
 twistor prescriptions for super Yang-Mills}, {\tt hep-th/0404085}.


\bibitem{Zhu}
C.~J.~Zhu, {\it The googly amplitudes in gauge theory},  JHEP {\bf 0404}
 (2004) 032, {\tt hep-th/0403115}.

\bibitem{Georgiou:2004wu}
G.~Georgiou and V.~V.~Khoze, {\it Tree amplitudes in gauge theory
as scalar MHV diagrams}, JHEP {\bf 0405} (2004) 070, {\tt hep-th/0404072}.

\bibitem{WuZhutwo} J-B. Wu and C-J Zhu, {\it MHV Vertices and
Scattering Amplitudes in Gauge Theory}, {\tt hep-th/0406085}.

\bibitem{WuZhuthree}
J-B. Wu and C-J Zhu, {\it MHV Vertices and Fermionic Scattering
Amplitudes in Gauge Theory with Quarks and Gluinos}, {\tt hep-th/0406146}.

\bibitem{Bena:2004ry}
I.~Bena, Z.~Bern and D.~A.~Kosower, {\it Twistor-space recursive
formulation of gauge theory amplitudes}, {\tt hep-th/0406133.}

\bibitem{dk}
D.~Kosower, {\it Next-to-Maximal Helicity Violating Amplitudes in
Gauge Theory}, {\tt hep-th/0406175}.

\bibitem{ggk}
G.~Georgiou, E.~W.~N.~Glover  and V.~V.~Khoze,
{\it Non-MHV Tree Amplitudes in Gauge Theory}, JHEP {\bf 0407}, 048 (2004),
{\tt hep-th/0407027}.



\bibitem{pv}
G.~Passarino and M.~J.~G.~Veltman,
{\it One Loop Corrections For
E+ E- Annihilation Into Mu+ Mu- In The Weinberg
Model,}
Nucl.\ Phys.\ B {\bf 160}, 151 (1979).






\end{thebibliography}
\end{document}